\documentclass[
 aps,
 prb,
 amsmath,
 amssymb,
 citeautoscript,
 reprint,
 superscriptaddress,
 floatfix
]{revtex4-1}

\usepackage[english]{babel}
\usepackage[colorlinks=true,linkcolor=blue,citecolor = blue]{hyperref}
\usepackage{color,soul}
\usepackage[normalem]{ulem} 
\usepackage{lipsum}
\usepackage{graphicx}
\usepackage{float}
\usepackage[dvipsnames]{xcolor}
\usepackage{comment}

\begin{document}
\title{Electron cascades and secondary electron emission in graphene under energetic ion irradiation}

\author{Henrique V\'{a}zquez}
\affiliation{Helsinki Institute of Physics and Department of Physics, University of Helsinki, P.O. Box 43, 00014 Helsinki, Finland}

\author{Alina Kononov}
\affiliation{Department of Physics, University of Illinois at Urbana-Champaign, Urbana, Illinois 61801, USA}

\author{Andreas Kyritsakis}
\affiliation{Helsinki Institute of Physics and Department of Physics, University of Helsinki, P.O. Box 43, 00014 Helsinki, Finland}

\author{Nikita Medvedev}
\affiliation{Institute of Physics, Academy of Science of Czech Republic, Na Slovance 1999/2, 18221 Prague 8, Czechia}
\affiliation{Institute of Plasma Physics, Czech Academy of Sciences, Za Slovankou 3, 182 00 Prague 8, Czechia}

\author{Andr{\'e} Schleife}
\email{schleife@illinois.edu}
\affiliation{Department of Materials Science and Engineering, University of Illinois at Urbana-Champaign, Urbana, IL 61801, USA}
\affiliation{Materials Research Laboratory, University of Illinois at Urbana-Champaign, Urbana, IL 61801, USA}
\affiliation{National Center for Supercomputing Applications, University of Illinois at Urbana-Champaign, Urbana, IL 61801, USA}

\author{Flyura Djurabekova}
\email{flyura.djurabekova@helsinki.fi}
\affiliation{Helsinki Institute of Physics and Department of Physics, University of Helsinki, P.O. Box 43, 00014 Helsinki, Finland}

\begin{abstract}
Highly energetic ions traversing a two-dimensional material such as graphene produce strong electronic excitations.
Electrons excited to energy states above the work function can give rise to secondary electron emission, reducing the amount of energy that remains the graphene after the ion impact.
Electrons can either be emitted (kinetic energy transfer) or captured by the passing ion (potential energy transfer).
To elucidate this behavior that is absent in three-dimensional materials, we simulate the electron dynamics in graphene during the first femtoseconds after ion impact.
We employ two conceptually different computational methods: a Monte Carlo (MC) based one, where electrons are treated as classical particles, and time-dependent density functional theory (TDDFT), where electrons are described quantum-mechanically.
We observe that the linear dependence of electron emission on deposited energy, emerging from MC simulations, becomes sublinear and closer to the TDDFT values when the electrostatic interactions of emitted electrons with graphene are taken into account via complementary particle-in-cell simulations. 
Our TDDFT simulations show that the probability for electron capture decreases rapidly with increasing ion velocity, whereas secondary electron emission dominates in the high velocity regime.
We estimate that these processes reduce the amount of energy deposited in the graphene layer by 15\,\% to  65\,\%, depending on the ion and its velocity. This finding clearly shows that electron emission must be taken into consideration when modelling damage production in two-dimensional materials under ion irradiation.
\end{abstract}

\maketitle

\section{Introduction}

Two-dimensional (2D) materials promise a myriad of new applications such as ultracompact electronics\cite{wang2012electronics}, nanosensors of unprecedented performance\cite{yuan2013graphene,han2011steam}, and water desalinators\cite{liu2016two,you2016graphene,madauss2017fabrication}, among others.
These applications cannot be realized with traditional, oftentimes low-precision, manufacturing techniques and require new, high-precision tools for modification of 2D materials.
In this respect, swift heavy ions (SHI), i.e.\ ions heavier than carbon with energies above 100 keV per nucleon, were shown to modify materials on the nanometer scale\cite{lang2020fundamental}. This feature makes this type of irradiation promising for tailoring single-layer materials.

To date, there have been only few studies on the effects of SHI irradiation in 2D materials. For example, experiments with SHIs under grazing incidence showed appearance of micron-size defects in graphene \cite{Akcoltekin2011} and MoS$_2$ \cite{madauss2017defect}.  
Under normal incidence, the regions affected by ions are much smaller, on a few nanometer scale.

These defects can be identified in graphene with Raman spectroscopy \cite{Vaz16}. 
This technique is sensitive to changes in the bonding environment of carbon atoms; however, it is not capable of resolving the nature and structure of the defects.
Moreover, high reactivity of the induced defects in graphene with air molecules limits the use of {\it ex-situ} imaging techniques for accurate analysis of the damage size.

Atomistic simulations can be used to bypass these limitations of imaging techniques in 2D materials.
For example, molecular dynamics (MD) simulations of SHI irradiation of graphene suggested that SHIs produce pores in this single-layer material\cite{Vaz16}.
Moreover, the size of the simulated pores showed the same trend as the corresponding experimental Raman signal.
The same technique elucidated the role of a substrate in damage formation in the irradiated graphene\cite{Zhao2015} and the formation mechanism of catalytic sites in MoS$_{2}$ by SHIs\cite{madauss2018highly}.
These studies gave valuable insights into how 2D materials respond to SHI irradiation, although the models used to simulate SHI impact were based on approaches developed for bulk materials.
In order to improve the accuracy of theoretical predictions, it is necessary to understand and include surface-specific processes, that despite not being critical in bulk, might significantly affect the dynamics in 2D materials.

\begin{figure}
\includegraphics[width=0.9\columnwidth]{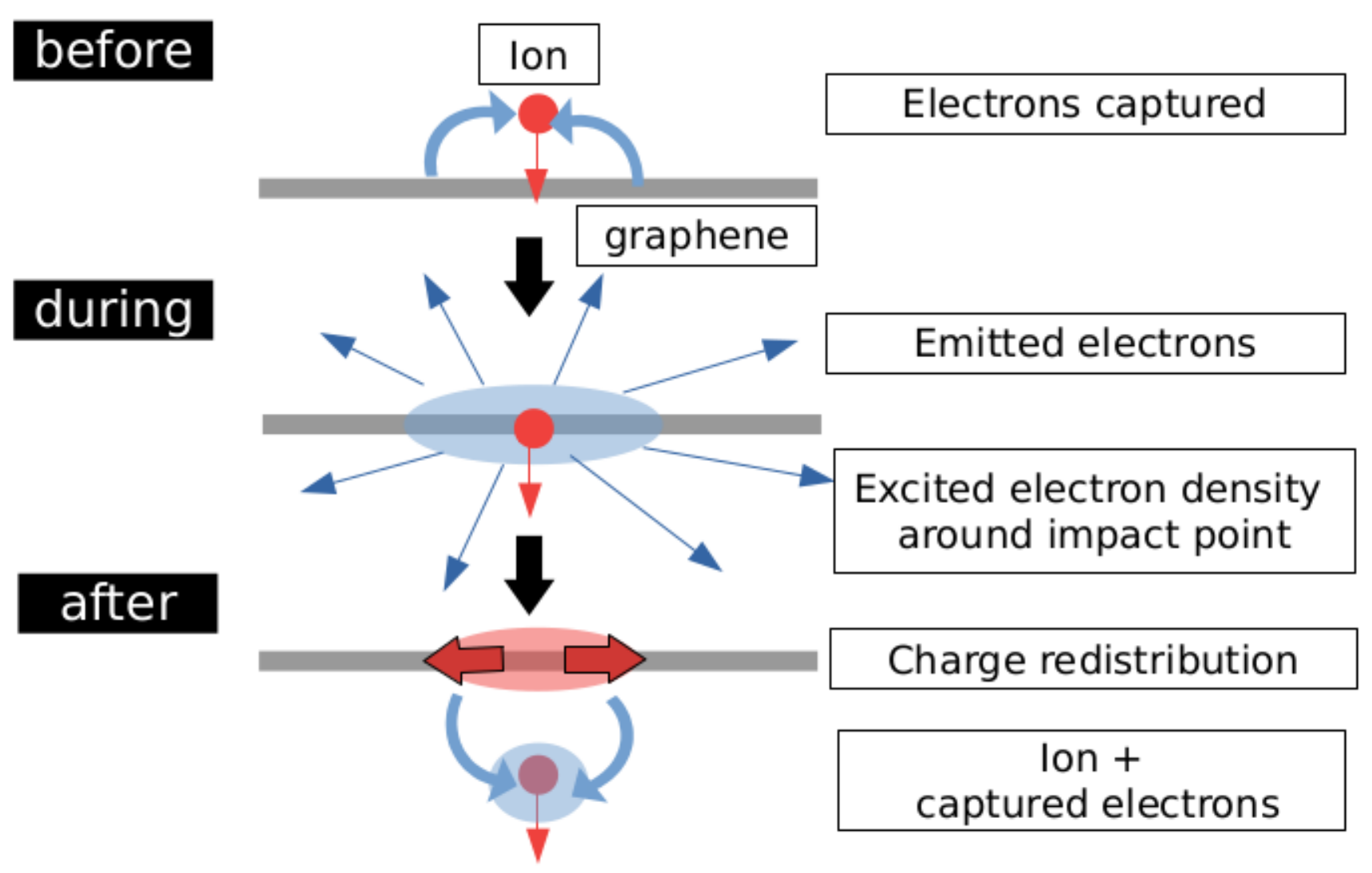}
\caption{\label{fig::schematics}
Schematics of the electron dynamics before (top), during (middle), and after (bottom) the impact of a SHI in graphene.
The approximate time scale of the dynamics shown here is on the order of femtoseconds.
}
\end{figure}

In bulk materials, highly energetic ions excite electrons along their trajectories, generating primary energetic $\delta$-electrons, which propagate outwards, exciting more electrons and generating an electronic cascade.
These excited electrons subsequently de-excite, transferring their energy to the atoms, which in turn may lead to defect formation in the irradiated material.
Despite similarities, there are several important mechanisms in 2D materials that are different from those in bulk (see schematics in Fig. \ref{fig::schematics}).
First, the excitation modes of 2D materials can differ from their bulk counterparts, which can influence energy deposition \cite{kononov:2020}.
Also, electrons excited to high energy states during a SHI impact may escape from the surface if their energy exceeds the barrier imposed by the work function.
This phenomenon is known as secondary electron emission (SEE).
Finally, a SHI captures electrons in bulk materials until its charge state reaches the equilibrium value, which happens within a few nanometers \cite{rozet1996etacha,imai2009equilibrium,osmani2011charge,Lee:2020}.
In 2D materials, the electron capture starts as the ion approaches the target, 
but this process may not have enough time to fully complete within the material and can become disrupted upon the SHI's exit from the target\cite{kononov:2020}.
Moreover, graphene might be especially efficient at supplying electrons;
Refs.\ \onlinecite{gruber2016ultrafast} and \onlinecite{schwestka2019charge} showed that during the impact of a highly charged ion (HCI) up to close to 30 and 70 electrons are captured and emitted respectively.

SEE and electron capture may play a major role in determining how the energy is deposited and redistributed after the ion impact in 2D materials, which may consequently affect defect formation.
Hence, in this work we aim to examine this hypothesis and quantify the loss of the initially deposited energy via these processes for different ion velocities and charge states.

To this end, we use two conceptually different simulation approaches: 
On one hand, we employ Monte Carlo based simulations of electron cascades, where the electrons are described as classical point-like charges.
On the other, we use time-dependent density functional theory (TDDFT) to describe quantum-mechanical behavior of electrons during the impact of an energetic ion.
Our Monte Carlo simulations predict as many as 100 emitted electrons during a single ion impact.
We interpret the large emission as an artifact of the simulation technique, which  does not include electrostatic interactions after emission.
When taking into account electrostatic interactions between emitted electrons and the charges in the layer via additional particle-in-cell simulations, we observe a significant reduction of SEE and a closer agreement with the values predicted by TDDFT.

From comparing the results of both types of simulations we improve the understanding of the primary mechanisms triggered in 2D materials by SHIs.
We show that SEE and electron capture carry away 15\,--\,65\,\% of the total energy that is initially deposited by the ion in the electronic subsystem. 
This reduction affects the size of structural defects arising from irradiation, which we expect to be smaller than what is obtained for 2D materials using bulk models that neglect SEE.

\section{Computational Methods}
\label{sec:methods}

Interaction of a SHI with the electronic subsystem is a complex, multi-scale phenomenon that involves processes such as electron-ion and electron-electron scattering, long-range Coulomb interactions, as well as the excitation and relaxation of electrons of the SHI and the target.
To describe the physics of these processes we use both classical and quantum-mechanical approaches.
Both of the methods have advantages and disadvantages in the context of SHI irradiation, which we discuss in short below.

Real-time TDDFT\cite{runge:1984,marques:2004,marques:2006,ullrich:2011,ullrich:2014} explicitly approximates the quantum-mechanical electron-electron interaction and includes the electron-ion Coulomb interaction.
It has recently been used to describe strong excitations in material surfaces created by either charged particles or electromagnetic radiation
\cite{wachter2012electron,ullrich1998electron,zhang2012ab,gruber2016ultrafast,kononov:2020}, and it is able to account for charge capture by the passing ion \cite{kononov:2020}.
In the context of this work, we use TDDFT to describe (i) energy deposition by the ion in the electronic system of the target, (ii) subsequent electron-electron scattering and electron-hole interactions in the excited state, and (iii) emission and capture of electrons.
However, the high computational cost of this method does not allow for simulations of very heavy and fast ions as well as large graphene sheets.
In particular, since simulations of fast ions over several femtoseconds require large simulation cells, we use TDDFT only to simulate ion impacts at relatively low velocities $\leq 4.7$ atomic units.

We employed asymptotic trajectory Monte Carlo simulations with the complex dielectric function\cite{medvedev2015time} (MC-CDF), where electrons are treated as point-like particles.
The physics described by MC-CDF relies on the explicit implementation of processes of interest and the choice of corresponding input parameters.
The asymptotic MC method has been successfully applied for decades to study electron cascades in bulk materials\cite{eckstein2013computer,gervais1994simulation,medvedev2010dynamics,akkerman2011ion} and was shown to provide a good approximation for high-velocity ions and high-energy electron scattering in various materials. 
The approximations used in this technique are less accurate near the Bohr velocity\cite{Garcia-Molina2012,medvedev2015time}.

This method was initially developed for bulk materials and its main disadvantage is that it does not include electrostatic forces between emitted electrons and positive charge left behind in the layer. During an ion impact, tens of electrons are emitted within a fraction of a femtosecond leaving positively charged holes in the graphene layer. Within this very short time interval, the created charge does not have time to equillibrate. Due to the lack of electrostatic interactions the method can not capture the rise of the emission barrier and consequent reduction of the SEE efficiency caused by the transient charge in the graphene layer.

In order to include the Coulomb interactions between electrons and holes, we couple the MC-CDF simulations with the particle-in-cell (PIC) method \cite{birdsall_plasma_1985,buneman_dissipation_1959,dawson_one-dimensional_1962}.
This ad hoc correction aims to describe the positive charge induced in the layer by emitted electrons and the corresponding increase of the barrier for electron emission due to Coulomb interactions. 
Only electrons that overcome the electrostatic barrier in PIC simulations are considered ``emitted''.
Similar electrostatic approaches have been used previously to model the barrier for electron emission in charged metallic clusters\cite{perdew1988energetics,kresin2008correlation,schone1994reintroduction,kalered2017work} and to study the dynamics of the emitted electrons in metals during laser irradiation\cite{zhou2014dynamics}.

In the following, we describe these techniques in detail.
Throughout this work, we use a$_0$ for the hydrogen Bohr radius and atomic units (a.u.) for velocities, where \mbox{1 a.u.} is the electron velocity in the first Bohr orbit.
An approximate conversion from velocity in a.u.\ to energy in MeV can be made through the expression $E \simeq 0.05 \times \frac{m}{2}v^{2}$, where $m$ is the ion mass in Dalton units and $v$ the velocity in a.u.

\subsection{Time-dependent density functional theory}

We performed real-time time-dependent density functional theory simulations using the Qbox/Qb@ll code \cite{draeger:2018,draeger:2017} to propagate the time-dependent Kohn-Sham equations \cite{Peuckert:1978,zangwill:1980,runge:1984,marques:2004},
\begin{multline}
\label{eq:tdks}
{\rm i}\hbar\frac{\partial}{{\partial}t}\phi_i({\bf r},t) = \\
     \left\{
          -\frac{\hbar^2\nabla^2}{2m} + \hat V_\text{ext}(t)
        + \hat V_\text{s}{[n({\bf r}, t)]}
      \right\} \phi_i({\bf r},t),
\end{multline}
in real time.
Eq.\ \eqref{eq:tdks} governs the dynamics of the electronic system, where ${\bf r}$ describes the spatial coordinate of electrons at time $t$, $\phi_i({\bf r},t)$ are Kohn-Sham states representing single-particle orbitals, $\hat V_{\rm ext}(t)$ describes the external potential due to the ionic system, and $\hat V_{\rm s}[n(t)]$ includes the Hartree electron-electron interaction and the quantum-mechanical exchange-correlation potential as a functional of the electron density $n({\bf r},t)$.

We used a plane-wave cutoff energy of 100 Ry and the adiabatic local density approximation (ALDA) \cite{zangwill:1980,zangwill:1981} for exchange and correlation.
The time-dependent external potential $\hat V_\text{ext}(t)$ is described by local and non-local parts of a pseudopotential, including the fast-moving projectile ion.
Explicitly describing all electrons in the system quickly becomes computationally prohibitive, and instead we use an HSCV pseudopotential \cite{vanderbilt:1985} with four valence electrons per carbon atom to describe the electron-ion interaction.
Each projectile ion is also described by a pseudopotential (RRKJ\cite{rappe:1990} in the case of Si$^{+12}$ and HSCV\cite{vanderbilt:1985} otherwise), where any occupied core states are pseudized.
While core electrons cannot be excited in this approach, which can lead to an effective reduction of electronic stopping at high ion velocities \cite{schleife:2015,ullah2018core}, the effect should be negligible for the velocities studied here, which lie well below the $\sim$13.5 a.u. threshold velocity\cite{lim:2016} at which an incoming ion could excite electrons across the 370 eV energy gap between $1s$ and $2s$ electrons in carbon\cite{NIST,erickson:1977}.

Fully converged ground-state Kohn-Sham wavefunctions from density functional theory \cite{dreizler:1990} for graphene were used as the initial condition for real-time propagation.
In the ground state calculation, the atomic forces were relaxed to less than 2 meV/{\AA}. Large simulation cells containing 112 carbon atoms and 150 a$_0$ (400 a$_0$) vacuum were needed to converge total charge transfer, including SEE and charge capture, to within 4\,\% for projectile ions with velocities of $v<2$ a.u. (2 a.u.\ $<v<5$\ a.u).
H$^{+}$, He$^{2+}$, Si$^{4+}$, Si$^{12+}$, and Xe$^{8+}$ were used as projectile ions in our simulations.

The Enforced Time Reversal Symmetry (ETRS) integrator \cite{castro:2004,draeger:2017} with a time step of 1 atto-second was used to evolve time-dependent Kohn-Sham equations, Eq.\ \eqref{eq:tdks}, for the electronic system because of its exceptional numerical accuracy for simulations of extended systems over thousands of time steps \cite{kang:2019,kononov:2020}.
In the beginning of the time-dependent simulation, each projectile ion starts 25 a$_0$ away from the graphene layer; it approaches and traverses the graphene at a constant velocity along a normal trajectory (see inset of Fig.\ \ref{fig::impactparam}).
Graphene nuclei are held at fixed positions because the few-fs time-scale of the simulations is too short for them to move appreciably.
As the projectile ion moves, we compute instantaneous electronic stopping $S(x)$ from Hellmann-Feynman forces acting on it.
Similar to the approach of Ref.\ \onlinecite{ojanpera2014electronic}, we then average $S(x)$ over the graphene thickness, taken as the inter-layer separation in graphite of 6.33 a$_0$\cite{delhaes2000graphite}.
The energy deposited in the graphene is simply given by the product of this average stopping power and the layer thickness.

SEE yields are determined by integrating the electron density $n({\bf r},t)$ over the volume outside the graphene and subtracting the number of electrons captured by the projectile ion, which is calculated using the orbital fitting technique described in Ref.\ \onlinecite{kononov:2020}.
Since the electron density decreases exponentially away from the graphene sheet into vacuum, there is no well-defined boundary between graphene and vacuum.
Here, we define the region outside 10.5 a$_0$ on either side of the graphene plane as outside of graphene.
With this choice, only $5\times 10^{-6}$ electrons lie within the vacuum region initially.
These techniques often produce nonintenger values for electron emission and capture, which can be interpreted as expectation values.
Finally, we perform a time of flight analysis\cite{ullrich:2011} on the electron density to calculate kinetic energy spectra of emitted electrons.

We stop the time propagation when electrons emitted from each side of the graphene merge across the periodic boundary in order to avoid unphysical results.
Thus, the computational limitations in cell size ultimately restrict the simulation length to a few fs.
Nevertheless, since the graphene charge plateaus within this time (see Fig.\ S7 of the Supplemental Material) and is converged with respect to vacuum size, this short simulation time is enough to capture the electron emission and electron capture processes in the material.

\subsection{\label{sec:MC}Monte Carlo simulations}

We also simulate the interaction of an incident ion with the target using a Monte Carlo model, which describes the propagation of individual particles according to the asymptotic trajectory event-by-event approach (see e.g.\ Ref.\ \onlinecite{jenkins2012monte}), as implemented in the TREKIS code.
All details of the model assumptions, cross section parameters, and numerical aspects of TREKIS were thoroughly described in Refs.\ \onlinecite{medvedev2015time,rymzhanov2016effects}.
The surface barrier model for electron emission used in the code was presented in Ref.\ \onlinecite{rymzhanov2015electron}, together with benchmarks and comparisons with available data for bulk materials.
These simulations are predictive, since all parameters are determined a priori, based on experiments and ab-initio simulations \cite{medvedev2015time,rymzhanov2016effects}. According to the Bohr criterion \cite{sigmund2016progress}, however,
this is strictly valid only for projectile ions with velocities higher than $\frac{2 Z}{v} \le 1$.

Within the MC formalism, the target is assumed to be a homogeneous arrangement of atoms, and different types of 
moving particles (SHI, electrons, holes) travel inside it until they reach a sampled site of interaction. 
Their free flight distance $d$ is sampled according to the Poisson distribution
\begin{eqnarray}
d&=&-\lambda \ln(\gamma)\\
\lambda&=&1/(N_\mathrm{at}\sigma)
\label{eq:flight}
\end{eqnarray}
where $\lambda$ is the mean free path, $\gamma$ is a random number within the interval $[0,1]$, $N_\mathrm{at}$ is the atomic density of the target, and $\sigma$ is the scattering cross section as discussed for the different particle types below.

The moving particles can interact with the atoms either elastically or inelastically. In elastic collisions, the particle transfers energy directly to the atoms, while in inelastic ones, it transfers energy to an electron. 
When an electron is excited, it leaves a hole in the core shell or in the valence band at the impact site.
The energy transferred in each inelastic collision is determined via an additional MC sampling step by evaluating the partial ionization cross sections for valence bands and core shells.
If ionization of the valence band was chosen, the energy level from which an electron is excited is determined according to the electron density of states of the target. If the ionization of a core shell is to take place, the energy level is chosen amongst the atomic core shells. The scattering event affects the energy and momentum of the moving particle, and the energy lost by it is divided between the created electron and hole, ensuring energy conservation. Momentum conservation determines the scattering angle of the new particles, while the azimuthal component of the momentum transfer direction is selected randomly.

In the SHI regime, the ions interact with materials mainly inelastically, depositing energy in the electronic subsystem, whereas the probability of collision with the target atoms is negligibly small.
Hence, in our model we focus only on the inelastic collisions of ions.
Electrons generated via inelastic scattering of the ions start their own trajectories. These electrons can interact both elastically and inelastically inside the material and might create more electron-hole pairs. Both types of interactions are taken into account in our simulations according to their respective cross sections. Holes interact with the target atoms only elastically and do not excite electrons directly. Note that in this approach, all the three particle types are scattered only by target atoms and do not interact between themselves.

The scattering cross sections in Eq.\ \eqref{eq:flight} are derived from different models depending on the type of interaction.
In order to model elastic scattering of excited electrons and valence band holes, where the carrier transfers kinetic energy to an atom without exciting new electrons, we use the Mott scattering cross section with the modified Molier screening parameter \cite{jenkins2012monte}.
Inelastic scattering is modelled for SHI and electrons using linear response theory based on the complex dielectric function (CDF)  $\varepsilon(\omega,q)$.
This formalism accounts for collective effects within the electronic system of the target, beyond the atomic approximation \cite{medvedev2015time,rymzhanov2016effects}.
The inelastic-scattering differential cross section is determined by the expression
\begin{equation}
\label{diff_cross}
\frac{d^{2}\sigma}{d(\hbar \omega)d(\hbar q)} = \frac{2(q_\mathrm{eff}(\nu,Z)e)^{2}}{n\pi \hbar^{2} \nu^{2}}\frac{1}{\hbar q}\text{Im} \left[-\frac{1}{\varepsilon(\omega,q)}\right],
\end{equation}
where $\text{Im} \left[-\frac{1}{\varepsilon(\omega,q)}\right]$ is the loss function of the material, $\omega$ and $q$ are energy and momentum transfer to the excited electrons, and $v$ and $q_\mathrm{eff}$ are velocity and effective charge of the particle 
. The effective charge $q_\mathrm{eff}(\nu,Z)$ is given by the Barkas formula\cite{gervais1994simulation} for ions (see Eq.\ (S7) in the Supplemental Material)
, while it is set to $-1$ for electrons.

In addition to scattering elastically as described above, core holes created by inelastic scattering events can decay via either Auger or radiative emission processes.
These events are determined based on the relative characteristic times of both processes, taken from the 2017 Electron Photon Interaction Cross Sections (EPICS-2017) database \cite{Cullen2018}, which combines experiments and ab-initio simulations, wherever available.
Emitted Auger electrons are considered secondary electrons and modeled in the same way as described above.
We do not observe any photons created in our simulations.

After sampling a free flight distance and collision or decay event, we repeat this procedure and continue simulating the trajectory of each particle 
generated. Since the time frame of these simulations is too short for any significant change in the target structure, all cross sections remain constant throughout the simulation.
We stop tracking ions and electrons when they leave the target or the simulation box and holes are instead reflected at the target boundaries.
We also stop tracking electrons when their energy drops below a cut off energy, chosen here as the work function of graphene, 4.6 eV \cite{yu2009tuning}.
We simulate these electron cascades for 2 fs, which is sufficiently long to observe saturation of electron emission, which occurs within the first femtosecond after a SHI impact (see Fig.\ S7 in the Supplemental Material).
The MC simulations were repeated 1000 times for each ion to obtain statistically reliable results\cite{medvedev2015time,rymzhanov2016effects}. We note that the simulation results converged already after 100 runs, with the accuracy of extracted quantities improving only by 1\% after completion of all 1000 independent runs.

In our simulations, we used material-dependent ionization energies \cite{pierson2012handbook} as an input to compute the ionization cross sections.
The electron mass was set to the mass of a free electron and the hole mass was calculated from the density of states of graphite \cite{ooi2006density,rymzhanov2016effects}.
In addition, optical data for graphite \cite{diebold1988angle} was employed to construct the CDF\cite{medvedev2015time}.
The choice of graphite parameters was motivated by the limitations of the TREKIS code\cite{medvedev2015time,rymzhanov2016effects}, originally developed for bulk materials.
In the Supplemental Material (see Fig.\ S8) we show that our results do not change significantly when we use the A-A stacked graphite density of states (DOS), which is similar to the DOS of graphene,\cite{Vaz16} instead of the experimental graphite DOS.
Therefore, we chose to use the experimental DOS of graphite for consistency with the CDF.
The MC-CDF framework used here also relies on the 3D CDF formalism.
In Fig.\ S2 of the Supplemental Material we compare the behavior of the inelastic mean free path of electron-electron interactions based on the cross sections obtained with 2D and 3D formalisms, and we see that the inelastic mean free path is very similar in both cases. 

These MC-CDF simulations provide transient radial distributions of the density and energy of excited electrons, of holes in the valence band, and in the core shells.
They also provide the energy transferred to the atomic system of the target and the kinetic energy spectra of electrons emitted from the surface.
However, for moving particles with velocities lower than and up to approximately the Bohr velocity $v_{0}$, the method becomes unreliable since the CDF formalism is based on linear response theory, i.e.\ first order perturbation theory, which breaks down at low velocities \cite{Garcia-Molina2012}.
Due to this limitation, we performed the MC-CDF simulations for ions with initial velocity above 1.8 a.u.
We also note that inelastic electron scattering can be described accurately only at electron energies above $\sim$40 eV, where the first Born approximation applies\cite{rymzhanov2016effects}.

\subsection{Particle-in-cell simulations}\label{subsec:PIC}

We couple the MC-CDF simulations with the particle-in-cell (PIC) method \cite{birdsall_plasma_1985,buneman_dissipation_1959,dawson_one-dimensional_1962} to include the effect of long-range electrostatic electron-electron and electron-hole interactions on the behavior of the emitted electrons. 
The PIC simulations follow the evolution of the electric field around the impact point, which is generated by the electron emission. This field, in turn, affects self-consistently the trajectories of the electrons in vacuum.
Since magnetic fields are negligible, electrons are subject only to the electric component of the Lorentz force, which is calculated iteratively at each time step solving the Poisson's equation by a finite-element method (FEM). 

In PIC, electrons are treated as elements of a continuous fluid in the phase space, which is then described by hydrodynamics.
In hydrodynamics, the number of particles is generally too large to be simulated explicitly, therefore the fluids are described by particle density and velocity distribution. 
The phase space in PIC simulations is divided into superparticles (SPs), whose number must be large enough to describe accurately the density and velocity distributions of the electrons. In our simulations, we used $10^{5}$ SPs to model the electron dynamics.
Since the number of emitted electrons in the MC-CDF simulations is rather small, to use the hydrodynamic approach, we assigned to each SP only a fractional number of electrons, similarly to Ref. \onlinecite{veske2019dynamic}.
The weight of each SP was chosen to be $w_\mathrm{SP}$=$n_\mathrm{emit} \times 10^{-5}$, where $n_\mathrm{emit}$ is the number of emitted electrons.
The movement of electrons is then described using the trajectory of each SP, which is tracked by solving numerically Newton's equation of motion with a time step of $\Delta t$=0.001 fs.
This allows for total simulation times between 7.5 fs for heavy ions and up to 90 fs for lighter ions with the convergence criterion of 0.3\,\%. 
\begin{figure}
\includegraphics[width=0.9\columnwidth]{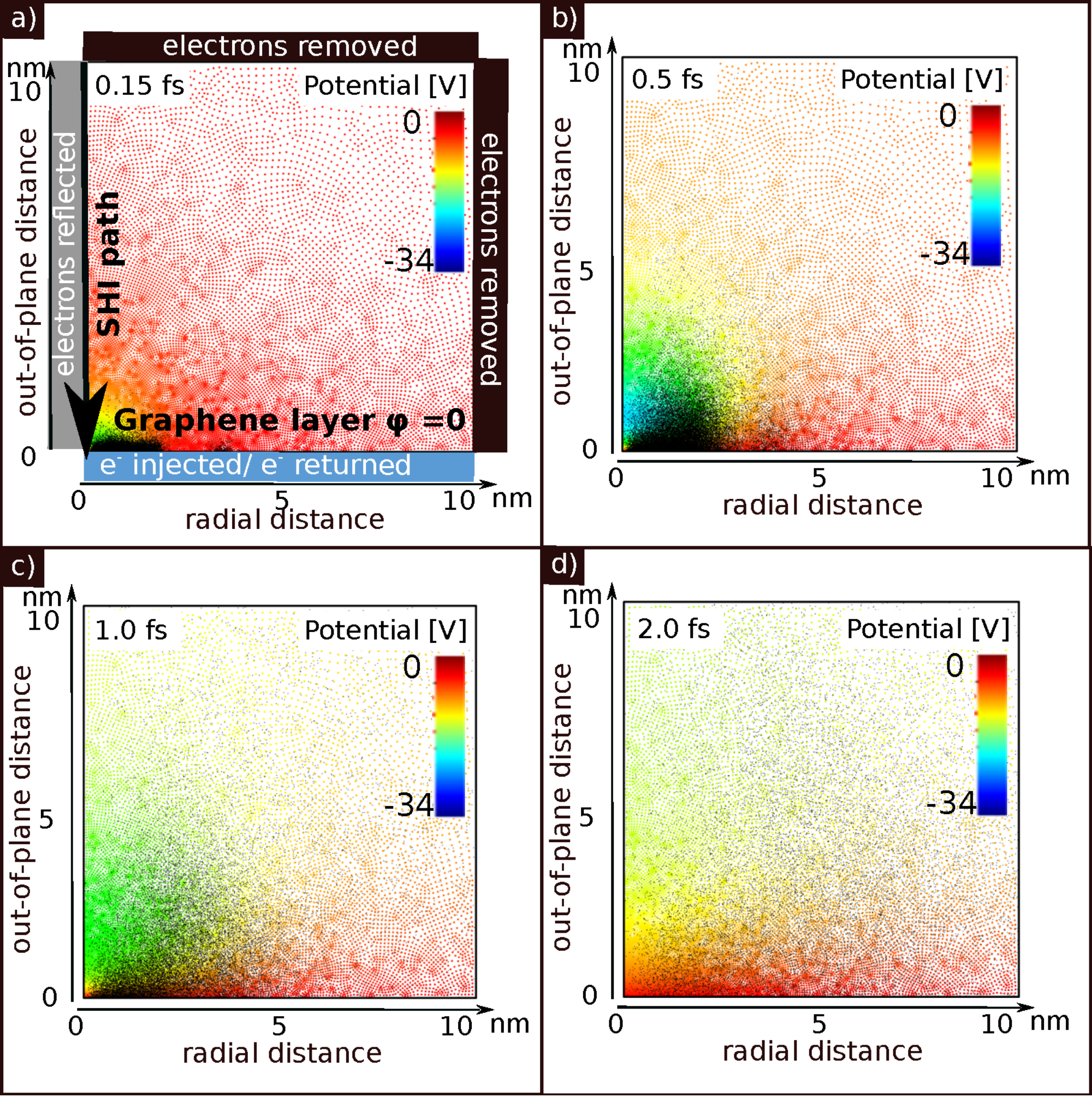}
\caption{\label{fig::PIC} 
Different snapshots from the femto-second real-time evolution, computed by 2D-PIC simulations, for emitted electrons from MC-CDF.
Top left panel includes the simulation setup and boundary conditions.
In each panel, the graphene sheet lies along the bottom edge and the ion travels along the left edge.
Black dots show superparticles which are used to track positions of electrons in vacuum.
The color coding shows the electrostatic potential generated by the emitted electrons.
}
\end{figure}

Due to rotational symmetry around the ion path, it is sufficient to simulate only a 2D radial cross section of the simulation space. Moreover, we can further decrease the size of the cell to only one quadrant by assuming symmetric SEE in up and down directions with respect to the graphene layer, as shown in Fig.\ \ref{fig::PIC}.
The symmetric geometry is motivated by the following rationale: The TREKIS code, which we employed for the MC-CDF simulations, gives the emitted energy spectra and the angular distribution of the emitted electrons as two separate outputs without giving the correlation between them.
Hence, for an emitted electron of given energy, we can not determine the direction of emission or if it leaves from the front or back surface.
Since we can not disentangle the contribution from forward and backward emission, we assumed instead symmetric SEE in both directions.
We use the aggregated energy spectra and angular distribution from both surfaces to generate the emitted electrons, but with only half the amount of the emitted electrons.
To obtain the total magnitude of the resulting SEE, we multiply the PIC SEE value by a factor of two.

We note that the simplified symmetric geometry is not fully accurate. In MC-CDF method, the SEE in the forward direction is 1.3\,--\,1.4 times higher. 
The difference of SEE in both directions calculated by TDDFT can differ by up to two to five times.
Hence, we performed additional PIC simulations where all the emission occurs from a single surface (front). The results, shown in Fig.\ S6 of the Supplemental Material, give us the upper bound of the electrostatic effect, since in these simulations the density of emitted electrons is much higher compared to the case of symmetric emission.

In this work we use a rectangular simulation box of $10\times10$ nm$^2$, as illustrated in the top left panel of Fig.\ \ref{fig::PIC}.
Increasing the cell size by a factor of 4 in each dimension only changes the emitted electron density and emitted energy by less than 5\,\% and 1\,\%, respectively. 
We also note that the element size in these finite-element simulations has a negligible influence of about 0.03\,\%.

In our simulations, we assume the graphene layer at the bottom of the simulation box ($z$=0) as a perfect conductor described by a Dirichlet boundary condition at zero electrostatic potential.
This approach implicitly models the positive charge left behind in the graphene as an image charge of every emitted electron.
Modeling graphene as a perfect conductor corresponds to approximating all the carriers in graphene as massless;
Gruber \textit{et al.}\ \cite{gruber2016ultrafast} reported very large electron currents in graphene during the passage of a HCI, which supports this approximation.

Electrons that reach the bottom graphene boundary (see Fig.\ \ref{fig::PIC}) are counted as ``recaptured'' by the material and no possible secondary cascades produced by these electrons are considered in our model.
The leftmost boundary ($r$=0) corresponds to the ion path and represents a rotational symmetry axis.
Electrons are reflected at this boundary.
The top and right boundaries in Fig.\ \ref{fig::PIC} assume no flux of electric field through them;
electrons crossing these are removed from the simulation and counted as ``emitted electrons.''

To represent emission of electrons, we inject a total of $N$ electron SPs at the bottom graphene layer at each time step $t_i$, with $N=\frac{1}{2} I(t_i) \Delta t / w_\mathrm{SP}$.
The emission rate $I(t)$ at time $t$ is obtained from the MC-CDF simulations (see details in Fig.\ S3 of the Supplemental Material), and the factor of $\frac{1}{2}$ is needed since we only explicitly simulate one of the surfaces of graphene.
Each SP is initialized at ($r$=$r_0$, $z$=0), with velocity ($v_r$, $v_z$), where $r_0$, $v_r$, $v_z$ are randomly selected according to the probability density distribution $P(r_0,v_r,v_z;t)$ obtained from our MC-CDF data.
We note that we used aggregate distributions, ignoring any correlation between energy and angle of the emitted electrons.
By collecting the exit statistics, we predict the fraction of electrons that are emitted far from the graphene sheet.
We perform these simulations using the efficient 2D-axisymmetric version of the FEMOCS framework \cite{kyritsakis2019atomistic,Veske2018279}, which has recently been extended to incorporate PIC \cite{veske2019dynamic}.

Representative PIC results for the electron dynamics in the vacuum above the graphene layer are depicted in Fig.\ \ref{fig::PIC}.
The four panels show the evolution in the sub-femtosecond range right after impact of a 91 MeV Xe ion.
The electron density and potential energy near the layer drop after 1 fs, when most of the electrons returned to graphene.
We also find that electron emission and return to the layer happen simultaneously and that most of the electrons do not travel further than 3\,--\,5 \AA~ from the layer.
This is illustrated in Fig.\ S4 in the Supplemental Material and we find that the rate of injection and return of electrons from PIC simulations (see Fig.\ S3 in the Supplemental Material) are almost perfectly superimposed.
After only 2 fs, the potential energy has dropped everywhere in the cell and no further evolution is observed.
This time scale agrees well with the time scale of electron emission observed in our TDDFT simulations (see Fig.\ S7 in the Supplemental Material).

The large fraction of returning electrons in the PIC simulations indicates that the electron energies are too low to overcome the electrostatic barrier of the positively charged graphene sheet.
By complementing the MC-CDF simulations with the PIC approach, we were able to imitate the effect of the transient change in graphene on the electron emission barrier due to the strong electrostatic field between the electrons and the image charge forming in the conducting graphene layer.
These mechanisms are implicitly captured by TDDFT and the ``returned" electrons can dissipate their energy by exciting electrons and/or plasmons in the graphene layer.

\section{Results and Discussion}

\subsection{Trajectory dependence}
\label{subsec:trajectory_dependence}

As described in Section \ref{sec:methods}, we employed two different approaches to simulate the impact of a SHI in graphene.
Differences between the two simulation techniques pose challenges for comparing results. For instance, in MC-CDF the projectile ion's trajectory is chosen randomly in graphene, which is represented as a random arrangement of atoms rather than an ordered atomic lattice. Conversely, in TDDFT the projectile ions are simulated in a deterministic fashion as they travel along a specific trajectory through the graphene crystal.
It was previously shown that the energy deposited by an ion in graphene depends on the impact parameter \cite{ojanpera2014electronic}. In the following, we further analyze the effect of the impact parameter on the number of emitted and captured electrons, which are the focus of the present study.

\begin{figure}
\includegraphics[width=0.9\columnwidth]{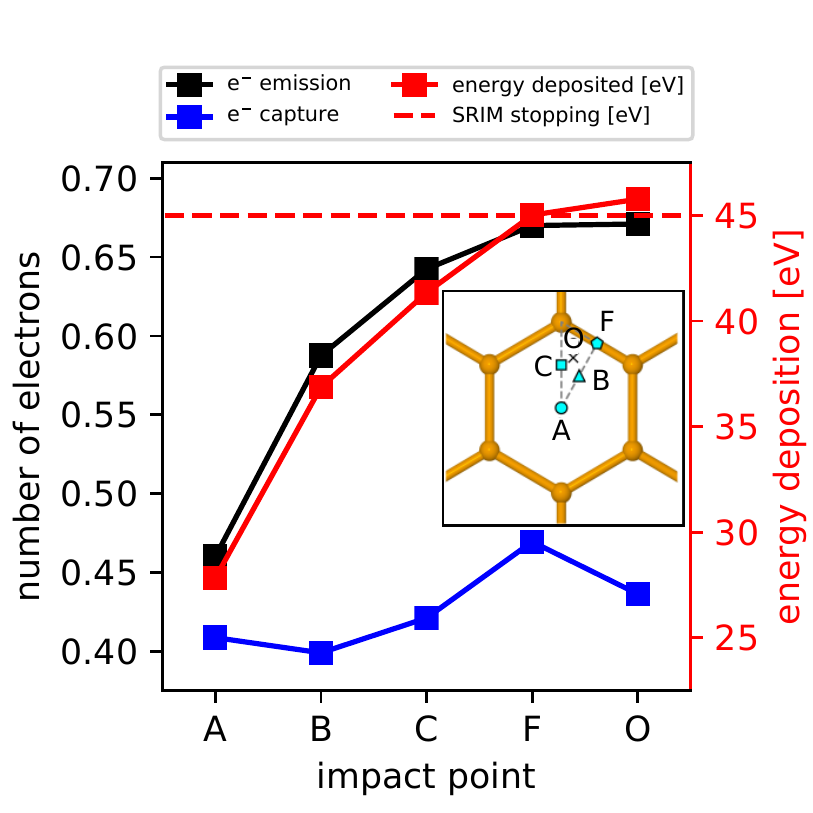}
\caption{\label{fig::impactparam}
Number of electrons emitted (black curve), number of electrons captured (blue curve), and energy deposited (red curve) by a 25 keV H$^{+}$ ion for different impact trajectories as calculated by TDDFT. 
The red dashed line shows the energy deposited by the same ion in a single layer (6.33 a$_0$ thick\cite{delhaes2000graphite}) of bulk graphite as calculated with SRIM\cite{ziegler2010srim}.
The inset shows the impact points (cyan) for the different ion trajectories normal to the graphene (orange); point O lies at the centroid of the gray dashed triangle.
}
\end{figure}

In Fig.\ \ref{fig::impactparam} we report the energy deposition, electron emission, and electron capture obtained by TDDFT for the \mbox{25 keV} H$^{+}$ ion with different impact trajectories, which are illustrated in the inset of Fig.\ \ref{fig::impactparam}. 
The deposited energy calculated by TDDFT along the O trajectory agrees well with SRIM's prediction, which is consistent with the TDDFT results reported in Refs.\ \onlinecite{krasheninnikov2007role,ojanpera2014electronic}. 
The highest and lowest electron emission and energy deposition correspond to the trajectories traversing the highest (F and O) and lowest (A) electron density, respectively. The difference in the number of emitted electrons between different trajectories can reach 50\,--\,60\,\%, while the deposited energy varies by up to 70\,\%.
Fig.\ \ref{fig::impactparam} also shows a clear correlation between SEE and the deposited energy for different trajectories, while the number of captured electrons depends on the trajectory only weakly.

In the remainder of this article, we present TDDFT results only for the most symmetric trajectory A due to the high computational cost of TDDFT simulations.
This corresponds to the impact position with the lowest electron density and the smallest energy deposition relative to the other impact parameters.

\subsection{Energy deposition}
\label{subsec:stopping_power}

\begin{figure}
\includegraphics[width=0.95\columnwidth]{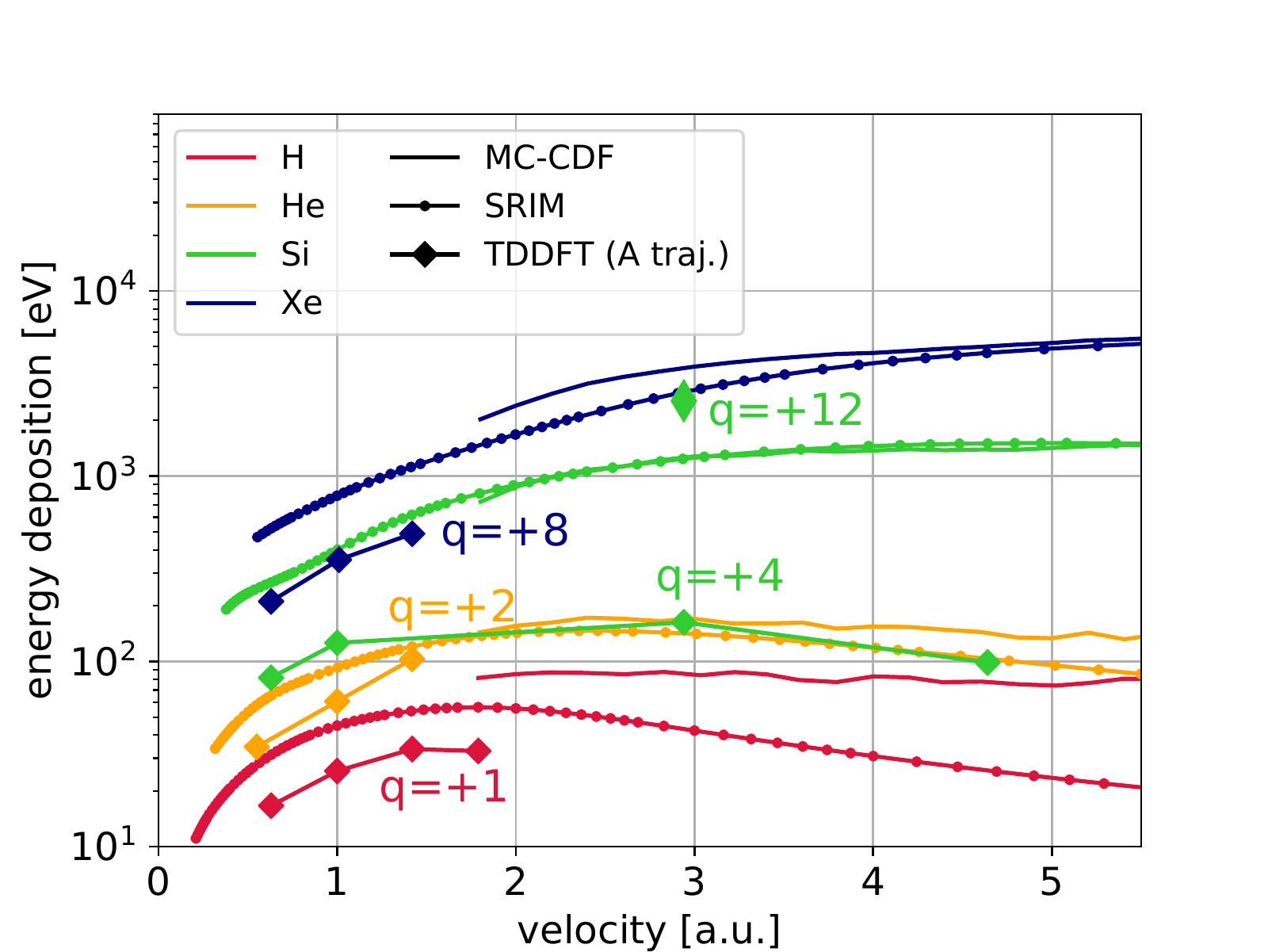}
\caption{Energy deposition in graphene for different projectile ion species and velocities. Results from TDDFT (diamonds), MC-CDF (solid curves), and SRIM (circles) are compared.
The text annotations correspond to the initial charge states of the ions in TDDFT; the MC-CDF calculations use an equilibrium effective charge given by the Barkas formula \cite{gervais1994simulation}. } \label{fig::stopping}
\end{figure}

In Fig.\ \ref{fig::stopping} we compare the energy deposited in graphene by ions of different type and velocity computed by TDDFT and MC-CDF with results from the SRIM database\cite{ziegler2010srim}.
We observe good agreement between the SRIM data and MC-CDF results, which is not surprising, considering that both models approximate graphene as a thin slice of bulk graphite. Moreover, MC-CDF employs the Barkas formula for the effective charge of projectile ions\cite{gervais1994simulation}, which was shown to give good agreement with SRIM \cite{medvedev2015time}.
While the agreement is particularly good for heavy ions such as Si and Xe (see Fig.\ \ref{fig::stopping}), MC-CDF overestimates the energy deposition for H ions, especially at high ion velocities. In this regime, the MC-CDF predictions are almost twice as large as the SRIM results.

In general, we observe that energy deposition in the TDDFT calculations is lower compared to the MC-CDF data (see Fig.\ \ref{fig::stopping}).
In some cases, this difference can be quite large, even more than an order of magnitude (see e.g.\ Si ions with velocity $>$ 2 a.u.). The discrepancy can be partially explained by the difference in the impact parameters used in the MC-CDF and TDDFT simulations:
As discussed in Section \ref{subsec:trajectory_dependence}, these TDDFT simulations were performed along trajectory A (see Fig.\ \ref{fig::impactparam} inset), which corresponds to the lowest electron density, while MC-CDF predictions represent an average over all possible impact parameters. 

In addition, the charge state of the projectile ion, which may change as the ion captures and loses electrons while traversing the graphene, affects the electronic stopping.
In TDDFT simulations, only the initial charge state is fixed and the charge dynamics are taken into account implicitly.
The ion velocities in this study, however, are rather high, and in most cases the ion does not spend sufficient time inside the layer to reach an equilibrium charge state.
Hence, the energy deposition still depends strongly on the initial charge state of the ion.
This is clearly illustrated in Fig.\ \ref{fig::stopping}, where we compare the energy deposition calculated in TDDFT for Si ions with 2.92 a.u. of velocity and initial charge states $q$=+4 and $q$=+12. We find that the energy deposition of the initially $q$=+12 ion is $\sim$20 times higher.

In the MC-CDF simulations, on the other hand, ions assume the equilibrium charge state (actual values shown in Fig.\ S9 of the Supplemental Material) from the beginning of the simulation.
As shown in Fig.\ \ref{fig::stopping}, the energy deposition obtained in MC-CDF simulations for Si with \mbox{2.92 a.u.} of velocity (effective charge state 5.2) falls between the values obtained in TDDFT for the two different initial charge states. 
We note, however, that the concept of effective charge state is employed as an ansatz in MC-CDF and other linear models\cite{lifschitz2004effective,maynard2000density} in order to reproduce accurately the stopping power measured experimentally in bulk materials. Therefore, the value of the effective charge itself, does not have to necessarily reproduce the actual charge state of the moving ion inside of the material.

While similar TDDFT simulations have shown that highly charged ions such as Si$^{+12}$ do not equilibrate within a single layer of material\cite{Lee:2020}, the lack of experimental measurements of the energy deposition in graphene does not allow us to determine which of the model predictions is more accurate.

\subsection{Secondary-electron emission}
\subsubsection{Energy spectra of emitted electrons}
\begin{figure}
\includegraphics[width=0.95\columnwidth]{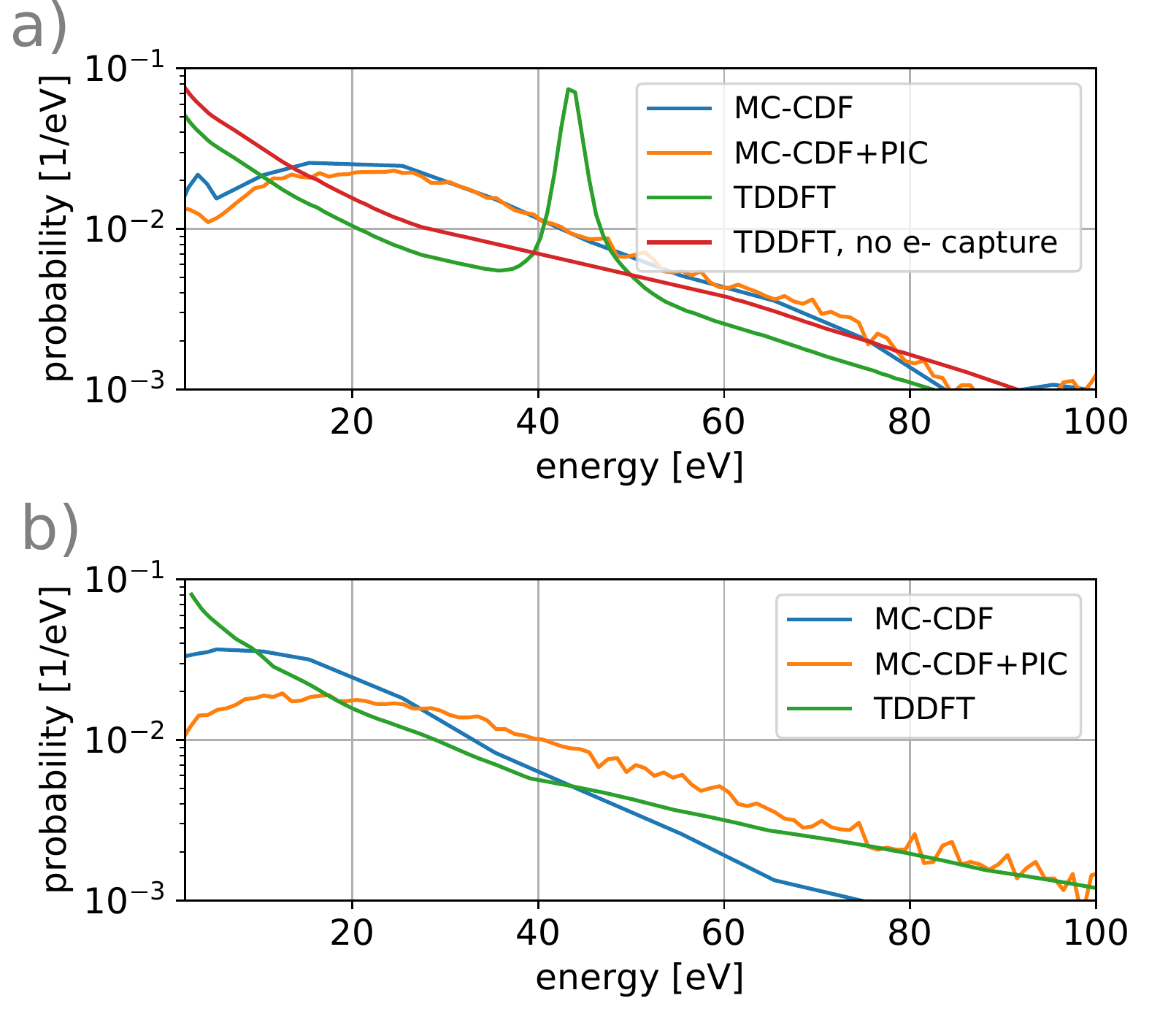}
\caption{\label{fig::SEE_spectrum} 
Normalized kinetic-energy spectra of the secondary electrons emitted from graphene after impact by a) H$^{+}$ 80 keV, b) Si$^{+4}$ 15 MeV.
TDDFT with (green) and without (red) captured electrons is compared to MC-CDF before (blue) and after (orange) including PIC simulations. MC-CDF results were computed with the Barkas effective charge.\cite{gervais1994simulation}
}
\end{figure}

In MC-CDF the kinetic-energy spectra of emitted secondary electrons can be obtained from the energy statistics of all emitted electrons. To extract this quantity in TDDFT, we instead apply a time-of-flight analysis\cite{ullrich:2011} to compute the spectrum from the time-dependent electron density.
We average the TDDFT spectra calculated for forward and backward emission in order to enable comparison with MC-CDF and MC-CDF+PIC.

The TDDFT spectrum for 80 keV H$^{+}$ in Fig.\ \ref{fig::SEE_spectrum}a features a prominent peak at \mbox{$\sim$44 eV}, which corresponds to the same electron velocity as that of the impacting proton.
A similar, though less pronounced peak at \mbox{$\sim$290 eV} lies beyond the scale shown in Fig.\ \ref{fig::SEE_spectrum}b for the 15 MeV Si$^{+4}$ ion.
The electrons constituting these peaks are largely localized around the ion, indicating that they have been captured by it.
Some of these electrons may be bound only weakly, e.g.\ in Rydberg or continuum states of the ion, and, hence, they could easily detach eventually\cite{Meckbach88}. 
Such weakly bound electrons which later detach into the vacuum would appear in experimentally measured SEE spectra; these are commonly referred to as convoy electrons \cite{brandt1977velocity}.
On the contrary, captured electrons that are more strongly bound to the ion would not be detected in measured SEE spectra. 

Since the simulation time of the TDDFT calculations is only a few fs, we cannot distinguish between ``convoy" and ``captured" electrons and, hence, we consider all electrons which left with the exiting ion ``captured".

Because MC-CDF does not include electron capture processes, this peak is not present in the energy spectrum calculated by this model. 

To enable consistent comparison between the MC-CDF and TDDFT results, we manually remove the peak between 32 and 60 eV from the TDDFT spectrum for the 80 keV H$^{+}$ ion, linearly connecting the probability values right before and after the peak and re-normalizing the distribution to unity.

We find that the normalized TDDFT and MC-CDF spectra compared in Figs.\ \ref{fig::SEE_spectrum}a and \ref{fig::SEE_spectrum}b follow a similar trend at high electron energies ($\gtrsim 40$ eV).
The agreement between the TDDFT and MC-CDF spectra for the \mbox{80 keV} proton further improves after the electron capture peak is removed from the TDDFT spectrum. 
Moreover, we see that the agreement between TDDFT and MC-CDF spectra for Si$^{+4}$ ions improves significantly for high energy SEE ($\gtrsim 60$ eV) when the MC-CDF method is coupled with PIC to include electron-electron and electron-hole electrostatic interactions for the emitted electrons. The same correction produces a less visible effect in the H$^+$ spectrum, which can be explained by the less efficient SEE for this ion (see Fig.\ \ref{fig::e_emission}a) and the corresponding reduction in the corrections introduced by PIC.

At low energies, below about 10\,--\,40 eV, however, the comparison between the spectra is poor. In this energy regime, the TDDFT spectra show a trend of increasing electron emission towards lower energies for both ions, whereas the MC-CDF and MC-CDF+PIC spectra exhibit a maximum at $<$ 20 eV.
The different behavior of the SEE spectra obtained with TDDFT and MC-CDF+PIC at low energies may be partially explained by the short time scales of the TDDFT simulations, which make it difficult to distinguish between low-energy emitted electrons and excess electrons in the vicinity of the graphene surface and could lead to overestimation of low-energy emission.
Moreover, the cross sections of inelastic electron scattering and inelastic mean free path (IMFP) adopted in the MC-CDF approach (see Fig.\ S2 in the Supplemental Material) are not sufficiently accurate at low energies\cite{rymzhanov2016effects} ($\lesssim 40$ eV), which may also affect the low-energy MC-CDF SEE spectra.
Finally, the electrostatic interactions between the impacting ion and electrons in the graphene are not included in either of our MC-CDF models, but they may also transiently reduce the emission barrier as the positively charged ion attracts electrons.

\subsubsection{SEE dependence on ion velocity and charge state} \label{subsub:see_ion_vel}

\begin{figure}
\includegraphics[width=0.95\columnwidth]{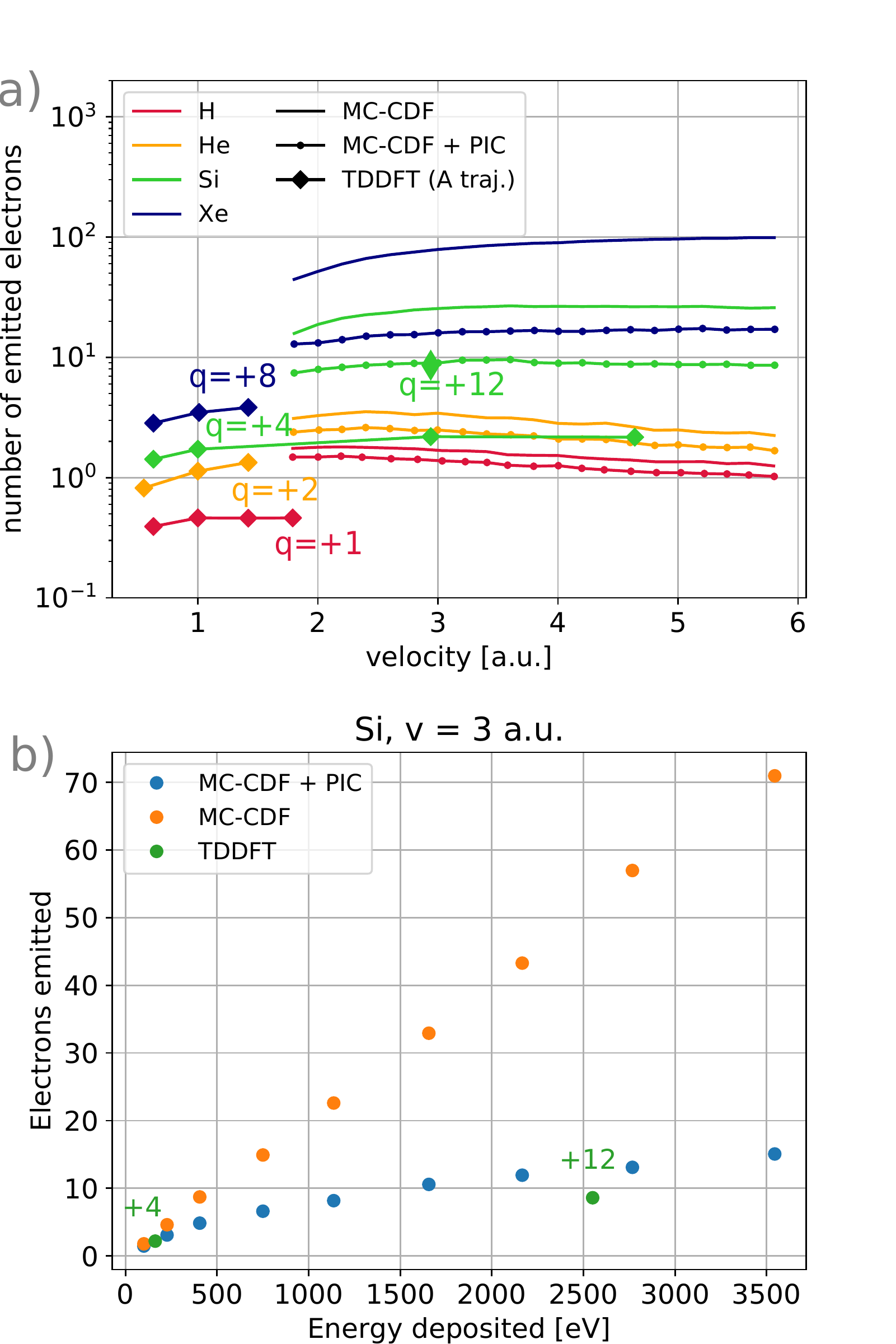}
\caption{\label{fig::e_emission}
a) Number of emitted secondary electrons in TDDFT (diamonds), MC-CDF (solid curves), and MC-CDF+PIC simulations (dots).
The TDDFT calculations used the initial charge states indicated beside the data points, while the MC-CDF(+PIC) calculations employed the Barkas effective charge\cite{gervais1994simulation}.
b) Electron emission vs.\ energy deposited in TDDFT (green), MC-CDF (orange), and MC-CDF+PIC simulations (blue) for Si projectile ions with $v$=2.92 a.u.
The TDDFT calculations used the initial charge states indicated beside the data points, while the MC-CDF(+PIC) calculations employed effective charge states in ascending order from +1 to +9.
}
\end{figure}

As discussed in Sec.\ \ref{subsec:stopping_power}, the ion species, velocity, and charge state affect the energy deposition (see Fig.\ \ref{fig::stopping}), and the energy deposition in turn is typically correlated with SEE\cite{rothard1990secondary}.
In the following, we analyze how these properties of the projectile ion affect the SEE process.
In Fig.\ \ref{fig::e_emission}a we show the number of emitted electrons for different impacting ions as calculated with the different methods.
We see that the total number of emitted electrons for a given ion  
velocity increases with the charge state of the ion in both TDDFT and MC-CDF.

The TDDFT results in the velocity range $v < 2$ a.u. show an increase of the number of emitted electrons with the ion velocity for all the studied ions except for H$^{+}$. 
The SEE of the latter peaks at \mbox{$v=1$ a.u.}
The results for the ions with \mbox{$v>2$ a.u.} were obtained only using MC-CDF and MC-CDF+PIC methods. These results show that both light ions, H$^{+}$ and He$^{+2}$  
emit less electrons with increasing velocity.
According to the MC-CDF+PIC results, the SEE of Si ions reaches the peak at \mbox{$v=3.5$ a.u.}, while no peak is observed for Xe ions, at least, in the range of velocities studied here. Based on these results, we infer that SEE does not increase monotonically as a function of ion velocity, but instead, at a certain velocity, which depends on the ion mass and charge state, SEE reaches its maximum and starts decreasing. The velocity corresponding to the maximal SEE seems to increase with the ion charge state.

\subsubsection{Comparison between TDDFT, MC-CDF and MC-CDF+PIC}

Due to the high computational costs associated with the large simulation cells required to model fast ions impacts, TDDFT calculations were mainly performed at low ion velocities.
The approximations underlying the MC-CDF model at these velocities are not valid, hence, the direct comparison of the results obtained with both methods is difficult.
To enable this comparison, we chose the Si$^{+4}$ ion and performed an additional set of TDDFT simulations for a wider range of velocities, which overlaps with the MC-CDF results.
The results presented in Fig. \ref{fig::e_emission}a show that the SEE predicted by TDDFT is much lower than that predicted by MC-CDF at the same velocities. Even though we did not perform similar high-velocity simulations for the rest of the ions, we see that the SEE at the highest studied velocity in TDDFT is much lower than the SEE obtained for the lowest possible velocity in the MC-CDF method. This clearly indicates that MC-CDF method overestimates efficiency of SEE compared to TDDFT.

In the following, we quantitatively discuss two possible reasons for the discrepancy between the MC-CDF and TDDFT results:
(i) the charge state of the ion in MC-CDF is fixed to the effective equilibrium value in bulk, whereas in TDDFT it evolves dynamically as the ion traverses the layer;
(ii) the electrostatic interactions of all charged particles are implicitly taken into account in TDDFT, but not in the MC-CDF approach. 

Since the results in Sec.\ \ref{subsub:see_ion_vel} suggest that SEE depends on the charge state of the ion,
we performed an additional TDDFT simulation for a Si$^{+12}$ ion with $v$=2.93 a.u.\ and found much higher SEE than that produced by the same ion with the lower charge state, Si$^{+4}$. Yet, when we compare the SEE obtained with both methods, we see that the SEE produced by the ion with higher initial charge state Si$^{+12}$ is still below the result obtained with the MC-CDF approach only (compare the green diamonds and the green solid line in Fig.\ \ref{fig::e_emission}a).

As previously mentioned, this large discrepancy between methods might be due to the lack of electrostatic interactions in MC-CDF.
We approach this problem by coupling the MC-CDF model to PIC simulations as described in Section \ref{subsec:PIC}. This allows us to follow explicitly the dynamics of the emitted electron cloud (see Fig.\ \ref{fig::PIC} and Fig.\ S7 in the Supplemental Material) and its attractive interaction with the charge induced in the layer. We show that overall the SEE obtained by the combined MC-CDF+PIC method agrees with the TDDFT data better (see the dotted lines in Fig.\ \ref{fig::e_emission}a), although the MC-CDF+PIC SEE data points are still persistently higher than the TDDFT ones. The PIC simulations compensate for the lack of electrostatic effects only {\it a posteriori}, so the possible effects of electrostatic interactions on SEE before and during the ion impact are still missing. Moreover, the differences in impact parameter and ion charge states still affect the TDDFT and MC-CDF results and do not allow a one to one comparison between methods. 

The reduction of SEE achieved due to the use of PIC simulation is, however, remarkable. It is particularly visible for the heavier ions, such as Si and Xe. The number of emitted electrons in these simulations was reduced by a factor of two and five, respectively. Moreover, for the lighter ions the correction introduced by PIC is less significant. This may be explained by fewer number of emitted electrons by these ions, since for low SEE, the electrostatic correction introduced by PIC is negligible. 

\subsubsection{SEE dependence on energy deposition}

Surprisingly, the SEE calculated for Si$^{+12}$ with the velocity of 2.92 a.u.\ matches very well with the result of the combined MC-CDF+PIC method (see a single diamond data point marked as q=+12 in Fig.\ \ref{fig::e_emission}a). The close agreement of this result with MC-CDF+PIC is likely to be explained by the higher energy deposition value compared to Si$^{+4}$ ion, since as we saw in Fig.\ \ref{fig::impactparam}, TDDFT simulations seem to show a correlation between SEE and energy deposition.
Strong correlation between the stopping power and the number of emitted electrons has already been established for bulk materials \cite{rothard1990secondary,hasselkamp2006particle,ritzau1998electron,sternglass1957theory}, where secondary electron excitation and emission rates were found to be roughly proportional to stopping power.

In Fig.\ \ref{fig::stopping} we showed that the value of the energy deposition is sensitive to the charge state of the ion. 
We analyze in Fig.\ \ref{fig::e_emission}b the correlation between the number of emitted electrons and the energy deposited in the layer by the passing ion as calculated by all three methods. The data shown corresponds to the Si ion with $v$=2.93 a.u. 
The TDDFT data is the same as in Fig.\ \ref{fig::e_emission}a, however, we performed additional MC-CDF and MC-CDF+PIC simulations with fixed ion effective charge values between q=+9 and q=+1 in descending order.
Both impact parameter and charge state of ions affect the amount of energy deposited by the ion and, consequently, the electron emission. By analyzing the electron emission as a function of energy deposition directly, we can exclude the effect of both factors and compare the TDDFT and MC-CDF results, avoiding the  uncertainty that arises from the definition of impact parameter and ion charge state.

In Fig.\ \ref{fig::e_emission}b, we see that Si$^{+4}$ and Si$^{+12}$ in TDDFT deposit as much energy along the trajectory A (see Fig.\ \ref{fig::impactparam}) as the Si ions in MC-CDF+PIC with a random impact parameter and the reduced effective charge state: $\lesssim +2$ and $\lesssim +8$, respectively. At the same time, both TDDFT data points fit very well within the dependence of the SEE on the deposited energy, obtained in the MC-CDF+PIC simulations.

In Fig. \ref{fig::e_emission}b we observe a linear growth of SEE with the deposited energy as calculated in MC-CDF.
The SEE in these calculations reaches as high value as 70 electrons per ion in the studied range of ion charge states. However, this efficiency is significantly reduced after the correction is introduced by the PIC simulations (blue dots in the figure), and the emission values become much closer to those obtained in TDDFT (green dots).

Despite the better agreement, we see that the SEE for Si$^{+12}$ in TDDFT is still slightly lower than that calculated in MC-CDF+PIC for the same energy deposition values. We remind here that the electron emission in PIC was simulated as symmetric, i.e.\ equal in forward and backward direction. This simplification is compatible with MC-CDF observations, where SEE was only 30-40\% higher in the forward direction compared to backward; in TDDFT however, we observe the emission mainly in the forward direction.
As one can see in Fig. S6 of the Supplemental Material, the emission in MC-CDF+PIC is further reduced by up to 30\,\% when all electrons are assumed to leave graphene from the same surface. Therefore, the difference in SEE seen in Fig. \ref{fig::e_emission}b for Si$^{+12}$ may be explained by the different preferential emission direction in the models.

We observe in Fig.\ \ref{fig::e_emission}b a saturation tendency for SEE with increase of the deposited energy for both the TDDFT and the MC-CDF+PIC simulations, while it is not observed for the pure MC-CDF simulations. This result differs from the approximately linear behavior of SEE vs.\ deposited energy observed for bulk materials \cite{rothard1990secondary}. This behavior in bulk is explained by the linear proportionality between the stopping power and the generation rate of secondary electrons \cite{sternglass1957theory}. While the model works well for bulk materials, it does not include the electrostatic interactions between the emitted electrons and the material surface. These interactions explain the sublinear behavior of SEE with the deposited energy, which we observed in both TDDFT and MC-CDF+PIC simulations.

\subsection{Energy removal from the layer by SEE and electron capture}

\begin{figure}
\includegraphics[width=0.95\columnwidth]{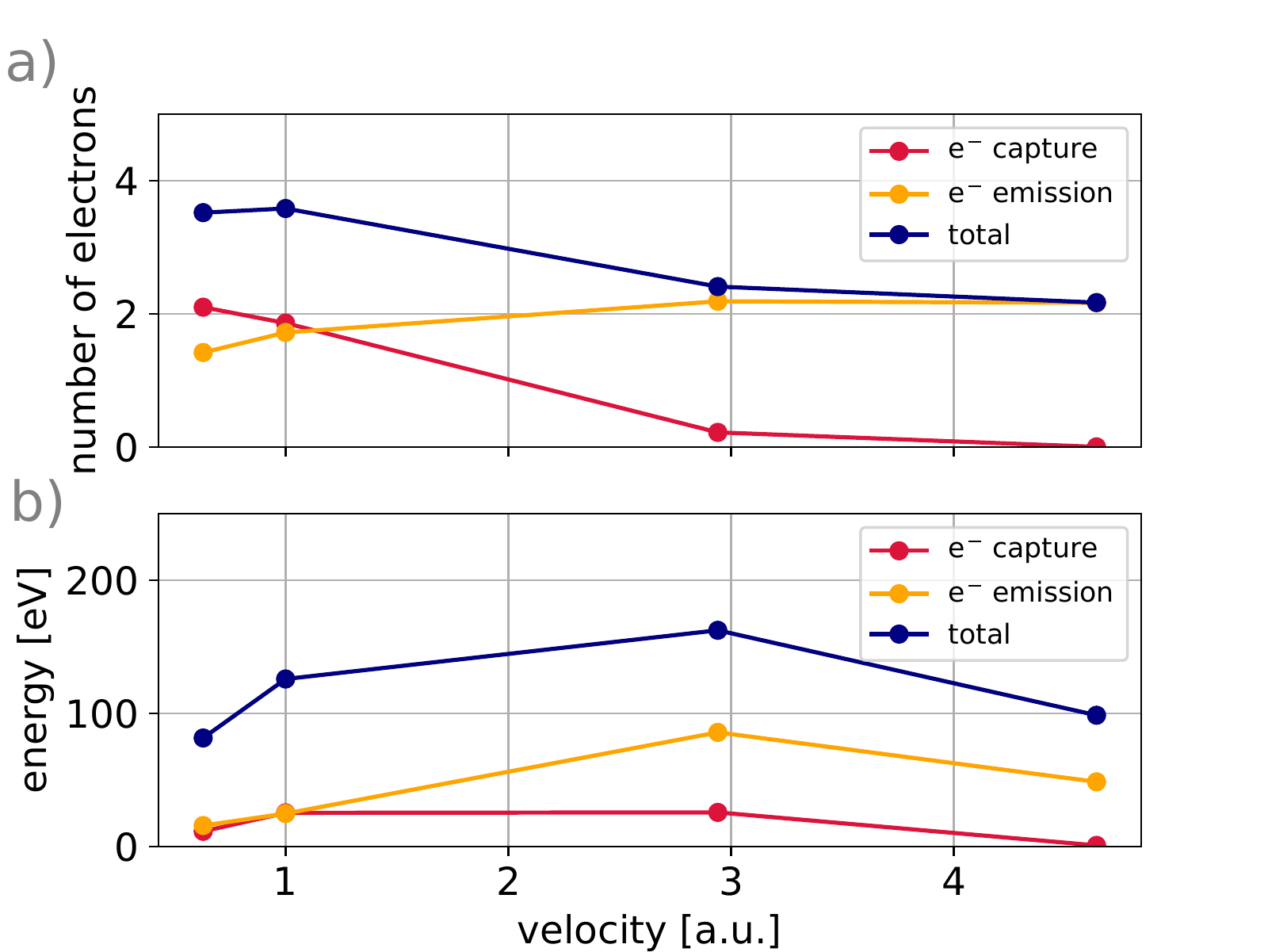}
\caption{\label{fig::emission_vs_capture}
a) Number of electrons emitted and captured in TDDFT for Si$^{+4}$ ions of varying velocities. b) Total energy deposited and dissipated via electron emission and electron capture for Si$^{+4}$ ions of varying velocities in TDDFT simulations.
}
\end{figure}

In this section we take a deeper look at the proportion of captured and emitted electrons, and we investigate the amount of energy carried away from the layer by each of these processes.
Since in TDDFT the electronic structure of the impacting ion and graphene enter the calculations directly, all electrostatic interactions between ions and electrons are also taken into account. By analyzing TDDFT calculations of the electronic excited states in the exiting ion, we are able to disentangle the number of captured electrons from the total number of emitted electrons.

In Fig.\ \ref{fig::emission_vs_capture}a we show separately the number of emitted and captured electrons as a function of ion velocity. The results show that electron capture dominates over electron emission at the ion velocities below 1 a.u.\ and drops rapidly to almost zero at 2.93 a.u., while the electron emission steadily increases with ion velocity.
However, we note that in the velocity range $0.5 \lesssim v \lesssim 1.8$ a.u., the numbers of emitted and captured  electrons are comparable.

We compute the energy loss for the quantum-mechanical electrons in TDDFT using the kinetic energy spectrum $P^{\mathrm{TDDFT}}(E)$ of the electrons outside the graphene layer. In these calculations both emitted and captured electrons are included. The kinetic energy spectrum is obtained from the time-of-flight analysis of the total electron density that escaped into vacuum.
The ``total" kinetic energy is then computed as
\begin{equation}
    E^\mathrm{kin}_{\mathrm{tot}}  = \int E\,P^{\mathrm{TDDFT}}(E)\,dE
    \label{eq:ekin}
\end{equation}
and subdivided into the kinetic energy of captured and emitted electrons, respectively, as
\begin{eqnarray}
E^\mathrm{kin}_{\mathrm{capt}} = & \frac{1}{2}m v^2 n_{\mathrm{capt}} \\
E^\mathrm{kin}_{\mathrm{emit}} = & E^\mathrm{kin}_{\mathrm{tot}} - E^\mathrm{kin}_{\mathrm{capt}},
\label{eq:ekin_parts}
\end{eqnarray}
where $v$ is the ion velocity and $n_{\mathrm{capt}}$ is the number of captured electrons.

In Fig.\ \ref{fig::emission_vs_capture}b, we show the TDDFT results for the total energy deposited by the Si$^{4+}$ ion (referred as "total" in the legend) and the kinetic energy that is taken away by the emitted and captured electrons separately, as given by \mbox{Eq.\ \eqref{eq:ekin}\,--\,\eqref{eq:ekin_parts}}.
The results show that in the velocity range below 1 a.u., i.e.\ when electron capture dominates, the combined energy loss from both processes is as high as 40\,\% of the total deposited energy.
With velocity increase, the electron capture becomes negligible and most of the energy is lost via electron emission.

\begin{figure}
\includegraphics[width=0.95\columnwidth]{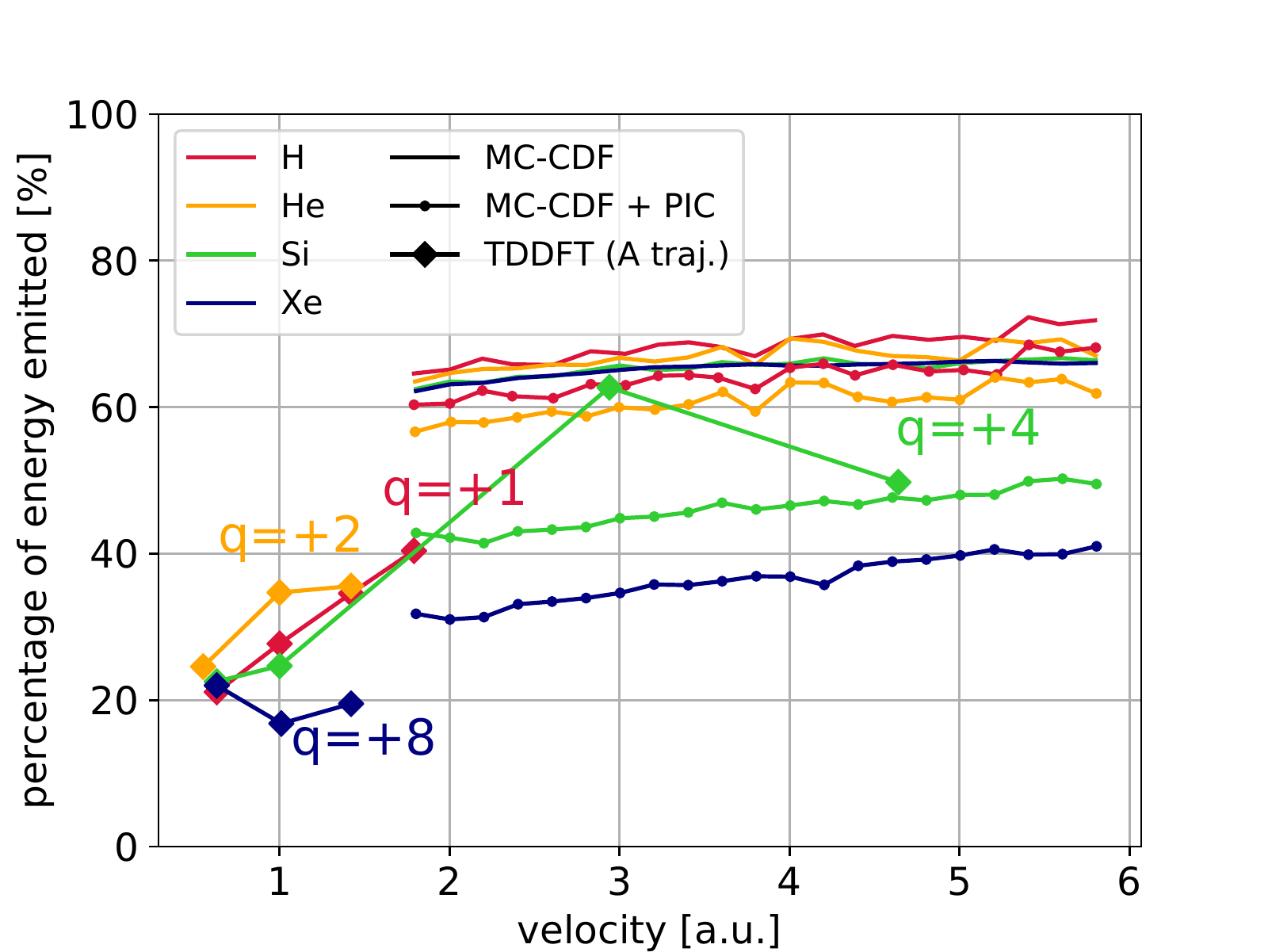}
\caption{Percentage of the initially deposited energy that is lost via electron emission in TDDFT (diamonds), MC-CDF (solid curves), and MC-CDF+PIC (points). The text annotations indicate the initial charge states of the ions in TDDFT; the MC-CDF calculations use the equilibrium effective charge state given by the Barkas formula\cite{gervais1994simulation}. }\label{fig::perc_emitted}

\end{figure}

Finally, we show in Fig. \ref{fig::perc_emitted} how much of the energy initially deposited by an ion into the graphene layer is lost in the electron emission processes as computed using TDDFT, MC-CDF, and MC-CDF+PIC. In this graph we see that the TDDFT simulations predict that \mbox{15\,--\,40\,\%} of the energy deposited by an ion with the velocity $\leq 1.8$ a.u.\ is subsequently emitted from graphene. 
In the high velocity range ($v\geq 1.8$ a.u.), the MC-CDF simulations show a deposited energy loss of up to 70\,\%. The PIC correction reduces significantly this percentage, at least, for heavy ions, however, in the velocity range ($v\simeq 1.8$ a.u.) where MC-CDF+PIC and TDDFT data should meet, the difference between both methods is still about 10-20\,\%. The only high velocity data points in TDDFT correspond to the Si$^{+4}$ ion. The energy emitted by this ion is as high as \mbox{45\,--\,60\,\%} of the initially deposited energy and is comparable to the MC-CDF+PIC predictions for the same ion.

Overall, Fig.\ \ref{fig::perc_emitted} shows that both TDDFT and MC-CDF+PIC predict that lighter ions with lower stopping powers cause greater fraction of energy loss via electron emission. This trend can be understood in terms of the electrostatic barrier for emission.
The higher energy deposited by higher charge ions is associated with a larger number of emitted electrons (see Fig.\ \ref{fig::e_emission}b), increasing the electrostatic barrier for electron emission.
Assuming similar energy distributions for the initially excited electrons, a lower percentage of excited electrons is then capable of overcoming the higher barrier and escaping the material after heavy ion impacts, leading to a lower percentage of initially deposited energy dissipating via electron emission.
This interpretation is further supported by the difference in behavior between MC-CDF and MC-CDF+PIC: the percentage of deposited energy lost to electron emission in MC-CDF is not sensitive to ion charge, indicating that the electrostatic interactions introduced by PIC are responsible for the stronger reductions in lost energy for heavy ions.

In summary, our results clearly show that electron emission and electron capture dramatically reduce the energy effectively deposited in graphene by the energetic ion. This reduction is expected to affect the size and morphology of the defects created by the ions and must be taken into account when modelling SHI impacts on single layer materials.

\section{Conclusions}

In this work we studied the electron dynamics triggered in graphene by swift heavy ions. We simulated the swift heavy ion impacts using both a quantum-mechanical approach, i.e. the time dependent density functional theory (TDDFT) method, and a classical approach by means of the Monte-Carlo method employing the complex dielectric function formalism (MC-CDF), where electrons are approximated as point-like particles.

Our Monte Carlo simulations resulted in a large number of emitted electrons. This number, however, was reduced significantly when electrostatic interactions between emitted electrons and holes in the graphene layer were taken into account via additional particle-in-cell (PIC) simulations.
The best agreement between methods was observed when we plotted the number of emitted electrons against the deposited energy. We observed a sublinear dependence of the electron emission on the energy deposition, which differs from roughly linear scaling reported previously for bulk materials.

The energy carried away by emitted electrons resulted in a 15\,--\,40\% reduction of the effective energy deposition in our TDDFT simulations at low velocities $v<1.8$ a.u.
At higher velocities $v>2$ a.u., this fraction increased to 40\,--\,70\%, which we observed in both TDDFT and MC-CDF+PIC simulations. Moreover, our simulations showed that light ions lose greater fraction of deposited energy to the emission processes than the heavy ions.

These findings suggest that defects created by swift heavy ions in 2D materials might be smaller than those created by the same ions in their bulk counterparts. We expect this reduction to be more pronounced for ions of low charge state and ion mass.

\begin{acknowledgments}
Henrique Va\'zquez thanks Alfredo Correa for his inspiring ideas and fruitful discussions.
Henrique Va\'zquez acknowledges support from the MATRENA doctoral programme.
This publication is partly based upon work from COST Action TUMIEE (CA17126), supported by COST (European Cooperation in Science and Technology), and partly based upon work supported by the National Science Foundation under Grant No.\ OAC-1740219. Andreas Kyritsakis was supported by the CERN K-contract (No. 47207461).
Partial financial support from the Czech Ministry of Education, Youth and Sports, Czech Republic (grants numbers LTT17015 and EF16\_013/0001552) is gratefully acknowledged by Nikita Medvedev.
Support from the IAEA F11020 CRP ``Ion Beam Induced Spatio-temporal Structural Evolution of Materials: Accelerators for a New Technology Era" is gratefully acknowledged.
Generous grants of computer time by CSC-IT are gratefully acknowledged.
This research is part of the Blue Waters sustained-petascale computing project, which is supported by the National Science Foundation (awards OCI-0725070 and ACI-1238993) and the state of Illinois.
Blue Waters is a joint effort of the University of Illinois at Urbana-Champaign and its National Center for Supercomputing Applications.
An award of computer time was provided by the Innovative and Novel Computational Impact on Theory and Experiment (INCITE) program.
This research used resources of the Argonne Leadership Computing Facility, which is a DOE Office of Science User Facility supported under Contract DE-AC02-06CH11357.
This work made use of the Illinois Campus Cluster, a computing resource that is operated by the Illinois Campus Cluster Program (ICCP) in conjunction with the National Center for Supercomputing Applications (NCSA) and which is supported by funds from the University of Illinois at Urbana-Champaign.
\end{acknowledgments}

\bibliographystyle{apsrev}
\bibliography{bibliography_CDF,bibliograhy_e_cascades,bibliography_tddft,bibliograhy1}
\end{document}


\title{Supplemental Material: Electron cascades and secondary electron emission in graphene under energetic ion irradiation}

\author{Henrique V\'{a}zquez}
\affiliation{Helsinki Institute of Physics and Department of Physics, University of Helsinki, P.O. Box 43, 00014 Helsinki, Finland}

\author{Alina Kononov}
\affiliation{Department of Physics, University of Illinois at Urbana-Champaign, Urbana, Illinois 61801, USA}

\author{Andreas Kyritsakis}
\affiliation{Helsinki Institute of Physics and Department of Physics, University of Helsinki, P.O. Box 43, 00014 Helsinki, Finland}

\author{Nikita Medvedev}
\affiliation{Institute of Physics, Academy of Science of Czech Republic, Na Slovance 1999/2, 18221 Prague 8, Czechia}
\affiliation{Institute of Plasma Physics, Czech Academy of Sciences, Za Slovankou 3, 182 00 Prague 8, Czechia}

\author{Andr{\'e} Schleife}
\affiliation{Department of Materials Science and Engineering, University of Illinois at Urbana-Champaign, Urbana, IL 61801, USA}
\affiliation{Materials Research Laboratory, University of Illinois at Urbana-Champaign, Urbana, IL 61801, USA}
\affiliation{National Center for Supercomputing Applications, University of Illinois at Urbana-Champaign, Urbana, IL 61801, USA}
\email{schleife@illinois.edu}

\author{Flyura Djurabekova}
\affiliation{Helsinki Institute of Physics and Department of Physics, University of Helsinki, P.O. Box 43, 00014 Helsinki, Finland}
\email{flyura.djurabekova@helsinki.fi}

\maketitle

\renewcommand\thefigure{S\arabic{figure}}
\renewcommand{\theHfigure}{S\thefigure}
\setcounter{figure}{0}

\renewcommand\thesection{S\Roman{section}}
\renewcommand\thetable{S\Roman{table}}
\renewcommand\theequation{S\arabic{equation}}

\section{Complex Dielectric Formalism}
\subsection{3D CDF oscillator fit} \label{app::CDF_fit}
TREKIS code \cite{medvedev2015time} employs the Complex Dielectric Function (CDF) formalism to calculate the cross sections. In this model, the inelastic cross sections are computed according to the equation (4) in the main text.

Within the Ritchie-Howie formalism, we decompose the Loss function $\text{Im} \left[\frac{-1}{\varepsilon(\omega,q)}\right]$ as a linear combination of oscillators (see \cite{ritchie1977electron}),
\begin{eqnarray}
\text{Im} \left[\frac{-1}{\varepsilon(\omega,q=0)}\right] = \sum_{i=1}^{N^{osc}} \frac{A_{i}\gamma_{i} \hbar \omega}{\left(\hbar^{2} \omega^{2} -E_{0i}^{2}\right)^{2}+(\gamma_{i} \hbar \omega)^{2}}
\label{eq::oscillator}
\end{eqnarray}

\begin{figure}
\includegraphics[width=\linewidth]{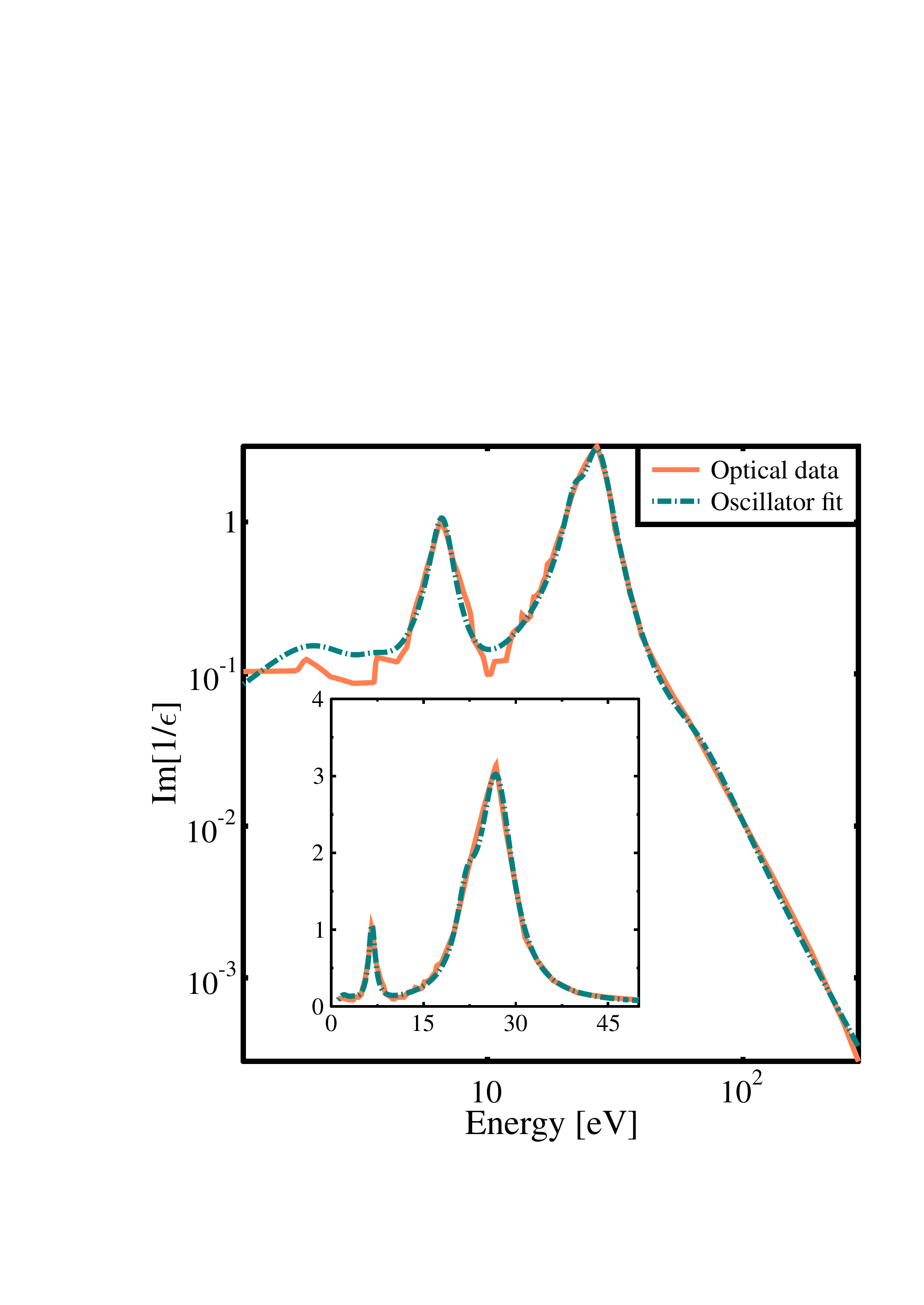}
\caption{Loss function from optical data (orange solid curve) and fit to the function with oscillators (green dashed curve). We show the data in both logarithmic scale (main figure) and in linear scale (inset).}\label{fig::oscill_plot}
\end{figure}

The coefficients of the oscillators can be fitted to experimental loss function data in the optical limit ($q=0$), and then extended into the plane $q>0$~\cite{ritchie1977electron}. In our case, we employed optical data of graphite\cite{diebold1988angle} for this purpose. In figure \ref{fig::oscill_plot} we show the original loss function together with the fitted oscillators; the oscillator parameters used can be found in table \ref{tab::oscill}.

\begin{table}[H]
\center
\begin{tabular}{| c | c | c | c |}
\hline
Oscillator & $A_{i}$ & $\gamma_{i}$ & $E_{i}$\\ \hline
1 & 75.74 &  51.85 &  63.96\\ \hline
2 & 516.14 &  6.66 &  27.01\\ \hline
3 & 76.61 &  4.03 &  22.10\\ \hline
4 & 8.70 &  1.32 &  6.65\\ \hline
5 & 0.20 &  1.65 &  3.66\\ \hline
6 & 0.55 &  2.25 &  2.23\\ \hline
\end{tabular}
\caption{Oscillator parameters fitted to the experimental loss function in graphite}
\label{tab::oscill}
\end{table}

\subsection{2D CDF Formalism} \label{app::2D_CDF}
The MC code TREKIS uses the 3D CDF formalism to calculate the cross sections. This approach was developed for isotropic three-dimensional materials and its application to a 2D material as graphene can lead to systematic errors in the electronic cross sections. In order to test this uncertainty, we compare the results of the 3D CDF with the 2D hydrodynamical model for electrons, developed initially by Fetter \cite{Fetter1973_2DCDF}. We employed the two fluids as proposed by Barton and Eberlein \cite{BartonEberlein1991_2DCDF} in order to treat graphene $\sigma$ and $\pi$ electrons separately. This model was shown to reproduce accurately the plasmon peaks in multilayer graphene \cite{Jovanovi2011_2DCDF} as well as in single layer graphene \cite{Jovanovi2011_2DCDF,Nelson2014_2DCDF}. Similarly to the 3D CDF formalism, in this model the loss function can be expressed as a sum of oscillators:
\begin{eqnarray}
\text{Im}\left[ \frac{-1}{\varepsilon(\vec{q}, \omega)} \right] & \simeq & V(q) \chi^{0}(\vec{q},\omega) \nonumber \\
&=& V(q) \sum_{\nu}\frac{n_{\nu}^{0} q^{2} / m^{*}_{\nu}}{s^{2}_{\nu}q^{2}+\omega_{\nu r}^{2} -\omega(\omega+i\gamma_{\nu})},
\end{eqnarray}
where $V(q) = \frac{e^{2}}{2 \varepsilon_{0} q}$ is the 2D Fourier transform of the Coulomb potential, $q$ is the 2D electron momentum and $m_{\nu}^{*}$ is the effective mass of the $\pi$ and $\omega$ electrons. We use the same oscillator parameters as in \cite{Jovanovi2011_2DCDF} (see table \ref{tab::oscill_2D}).

\begin{table}
\center
\begin{tabular}{| c | c | c | c |}
\hline
Oscillator & $n^{0}_{\nu}$ [$\text{nm}^{-2}$] & $\gamma_{\nu}$ [$\text{eV}$] & $\hbar \omega_{\nu r}$  [$\text{eV}$]\\ \hline
$\pi$ & 38 & 2.45 & 4.08 \\ \hline
$\sigma$ & 115  & 2.72  & 13.06 \\ \hline
\end{tabular}
\caption{2D oscillator parameters from }
\label{tab::oscill_2D}
\end{table}

Following the derivations in \cite{Jovanovi2011_2DCDF,Nelson2014_2DCDF}, the loss function of an incident particle in a 2D material can be expressed as
\begin{eqnarray}
E_{\text{loss}} & = & \int^{E_{\text{max}}} d\omega \omega P(\omega) \\
 P(\omega, v) & = & \int d^{2}\textbf{q} \frac{1}{4 \pi \varepsilon_{0}}\frac{\left(Z e\right)^{2}}{\pi^{2}} \times \nonumber ;\\
& & \left(\frac{2 q v_{\perp}^{2}}{(q v_{\perp})^2+(\omega -\textbf{q} \cdot \textbf{v}_{\parallel})^2}\right)^2 \text{Im}\left[\frac{-1}{\varepsilon(\omega, \textbf{q})}\right],
\end{eqnarray}
where $P(\omega)$ is the probability for an incident particle of charge $Z$ with perpendicular and parallel velocity components $v_{\perp}$, $v_{\parallel}$ to deposit an energy $\omega$ via electron scattering.

The inelastic mean free path of an electron traversing perpendicularly graphene can be obtained as
\begin{eqnarray}
    \lambda^{-1}(v) & = & \frac{1}{d_{\text{gr}}} \int^{E_{\text{max}}} d\omega P(\omega, v) = \\
    & = & \frac{1}{d_{\text{gr}}} \int^{E_{\text{max}}} d\omega \int_{q_{-}}^{q^{+}} d\textbf{q} \frac{1}{4 \pi \varepsilon_{0}} \frac{e^{2}}{\pi v^{2}} \times \nonumber\\
& & \left(\frac{2 q}{q^2+(\frac{\omega}{v})^2}\right)^2 \text{Im}\left[\frac{-1}{\varepsilon(\omega, \textbf{q})}\right],
\end{eqnarray}
where $d_{gr}$ is the thickness of the graphene layer and $q_{\pm}$ are the maximum and minimum momentum transfer in an electron-electron scattering $q_{\pm} = \sqrt{2 m_{e}/\hbar^{2}}(\sqrt{E} \pm \sqrt{E -\hbar \omega})$.

\subsection{Inelastic Mean Free Path comparison}

\begin{figure}[H]
\begin{center}
\includegraphics[width=0.9\linewidth]{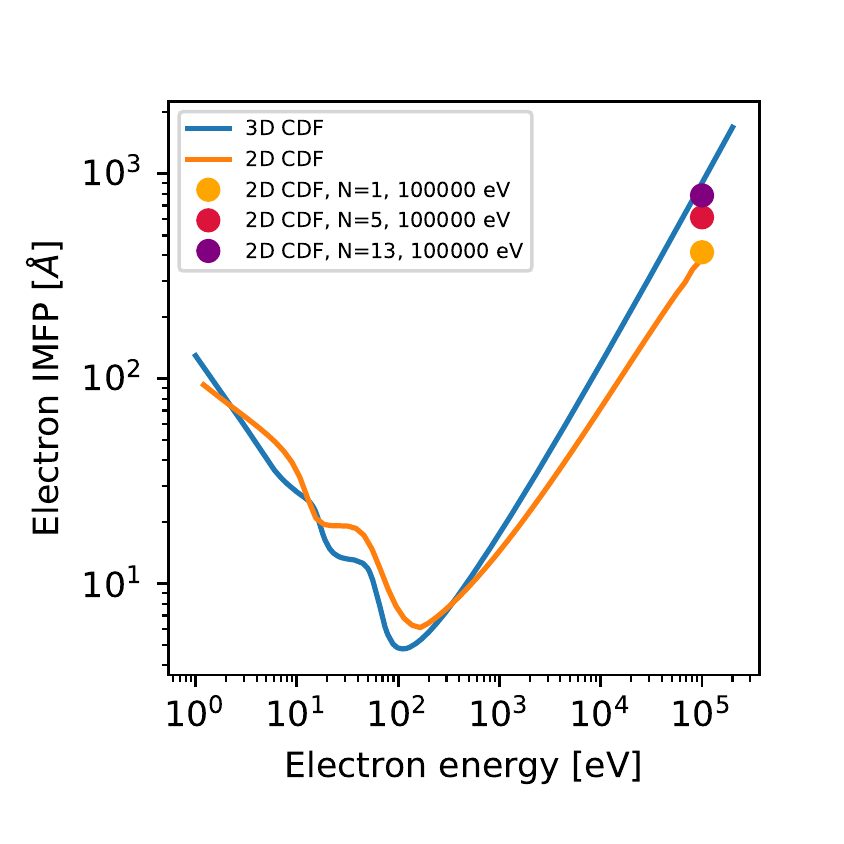}
\caption{Inelastic mean free path in graphene as calculated with the 2D CDF formalism (yellow curve) and with the 3D CDF (blue curve). The data points correspond to the IMFP in multilayer graphene for 100 keV electrons from Ref.\  \onlinecite{Jovanovi2011_2DCDF}. }
\label{fig:IMFP}
\end{center}
\end{figure}

The cross sections used in the MC code were calculated according to the 3D CDF formalism with optical data from graphite as input. In figure \ref{fig:IMFP} we compare the inelastic mean free path (IMFP) calculated with the 3D formalism and with the 2D formalism fitted to EELS data from graphene. 

We see that both approaches predict similar IMFP curves, nevertheless, the position of the peaks is not identical and the mean free path disagrees at some energies by up to a factor of two. Electrons with low kinetic energies exhibit a smaller IMFP in the 3D CDF model than in the 2D one. For electrons with energy higher than 200 eV this trend reverses, and the mean free path becomes lower in the 2D approach. We also obtain the IMFP for multilayer graphene using the interaction probability from\cite{Jovanovi2011_2DCDF} for electrons of energy 100 keV. We see that with increasing number of layers, the IMFP converges to the 3D CDF IMFP value. Therefore, we attribute the IMFP differences in both models mainly to the shift in plasmon peaks in the monolayer compared to bulk graphite.

\section{Particle-In-Cell Simulations}
\begin{figure}[H]
    \centering
    \includegraphics[width=\linewidth]{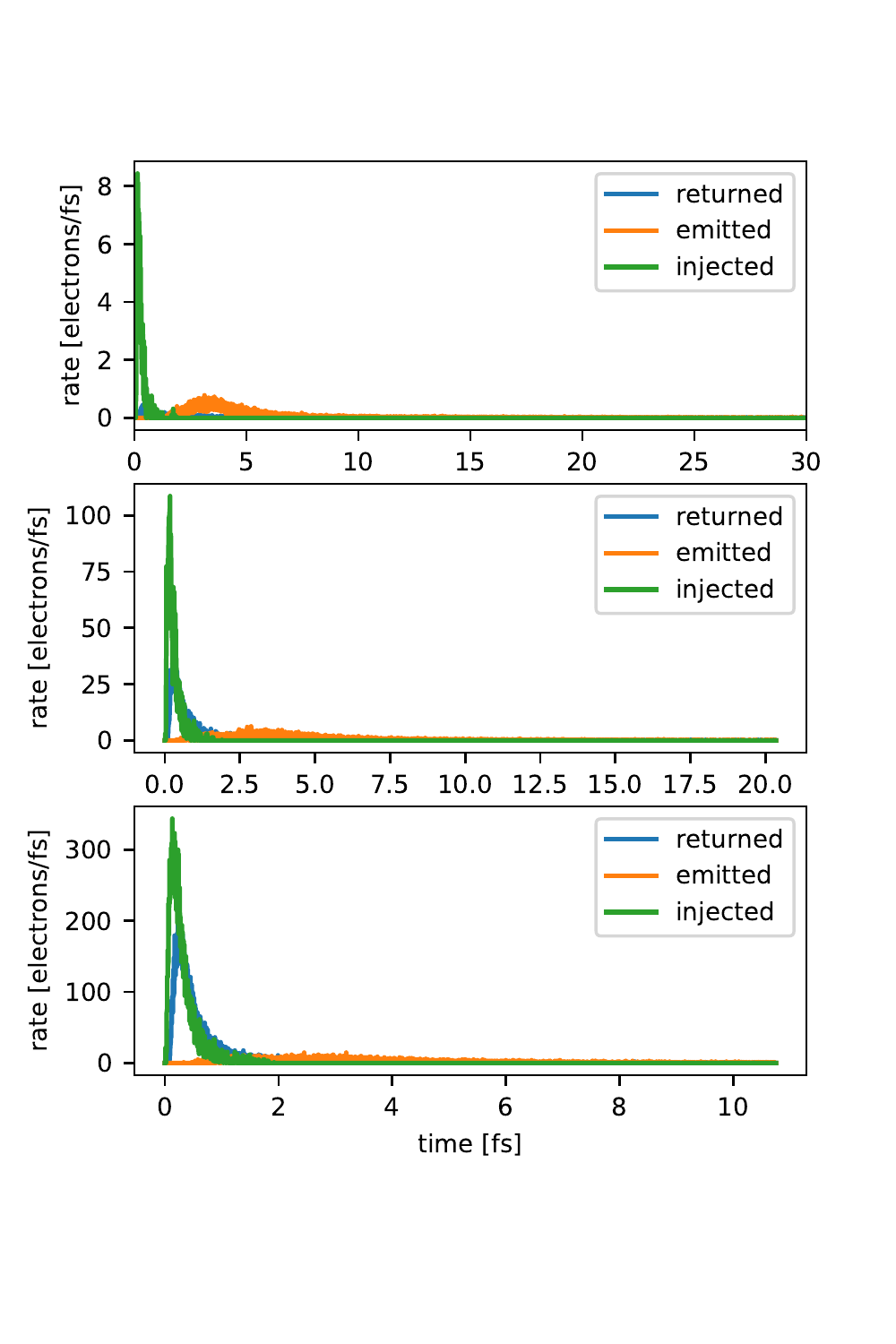}
    \caption{Time evolution of the rate of emission (orange curve), return (blue curve) and injection (green curve) of electrons in PIC simulations the ions H 80 keV (top graph), Si 15 MeV (middle) and Xe 91 MeV (bottom).
    }
    \label{fig:time_evolution}
\end{figure}

In figure \ref{fig:time_evolution} we show the evolution of the rate of emitted, returned and injected electrons for three ion impacts H 80 keV, Si 15 MeV and Xe 91 MeV in the PIC simulations. 
We see that for the injection and return rate are almost perfectly superimposed, what means that most of the electrons return immediately to the layer as they are emitted. We see that the proportion of returning electrons is much larger for the heavier ions Si 15 MeV and Xe 91 MeV than for H 80 keV. This is due to the small electrostatic potential generated by the lighter ions in the layer which deposit less energy. 

\vspace{1em}

\vspace{1em}

\begin{figure}[H]
    \centering
    \includegraphics[width=\linewidth]{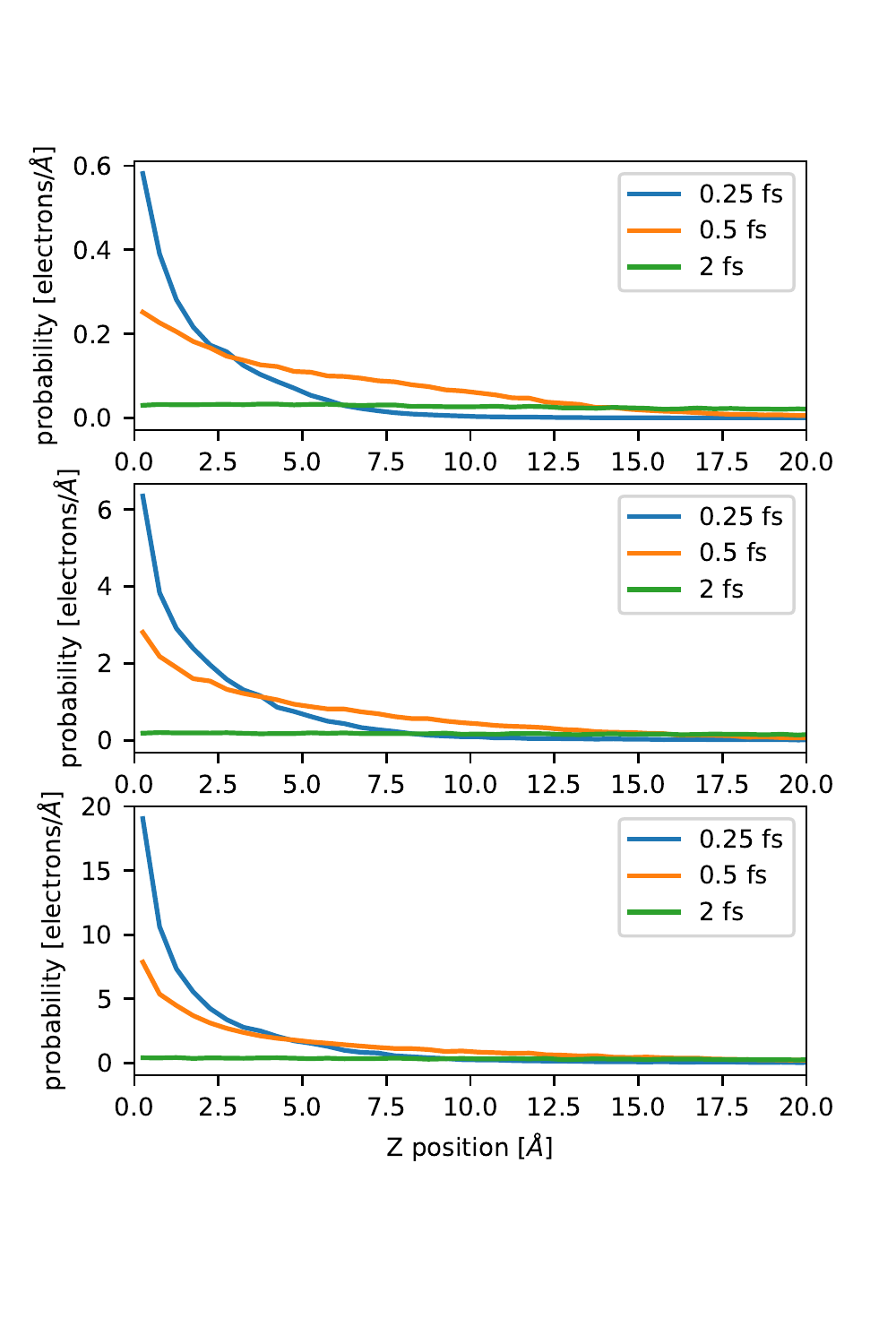}
    \caption{Position distribution in Z of the superparticles at times 0.25 fs (blue), 0.5 fs (orange) and 2 fs (green). We show these results for the ions H 80 keV (top), Si 15 MeV (middle) and Xe 91 MeV (bottom).
    }
    \label{fig:z_histogram}
\end{figure}

In Fig.\ \ref{fig:z_histogram} we show the Z-position distribution of the electron density at different times after the ion impact: 0.25 fs, 0.5 fs and 2 fs; we show the results for the ions H 80 keV, Si 15 MeV and Xe 91 MeV. We see that at 0.25 fs most of the charge density is confined between 1 \AA\ (Hydrogen) and 3 \AA\ (heavier ions) distance from the graphene layer. At 0.5 fs we already see a clear drop of the charge near the layer, and the density starts to spread outwards from the layer; at 2 fs we do not see anymore density near the layer, just a small tail of electron density already reaching the border of the cell. This shows that in the time span when the electron density is recaptured by the layer, the electron density had not time to travel far (between 0-5 \AA). 

\vspace{1em}

\begin{figure} [H]
    \centering
    \includegraphics[width=\linewidth]{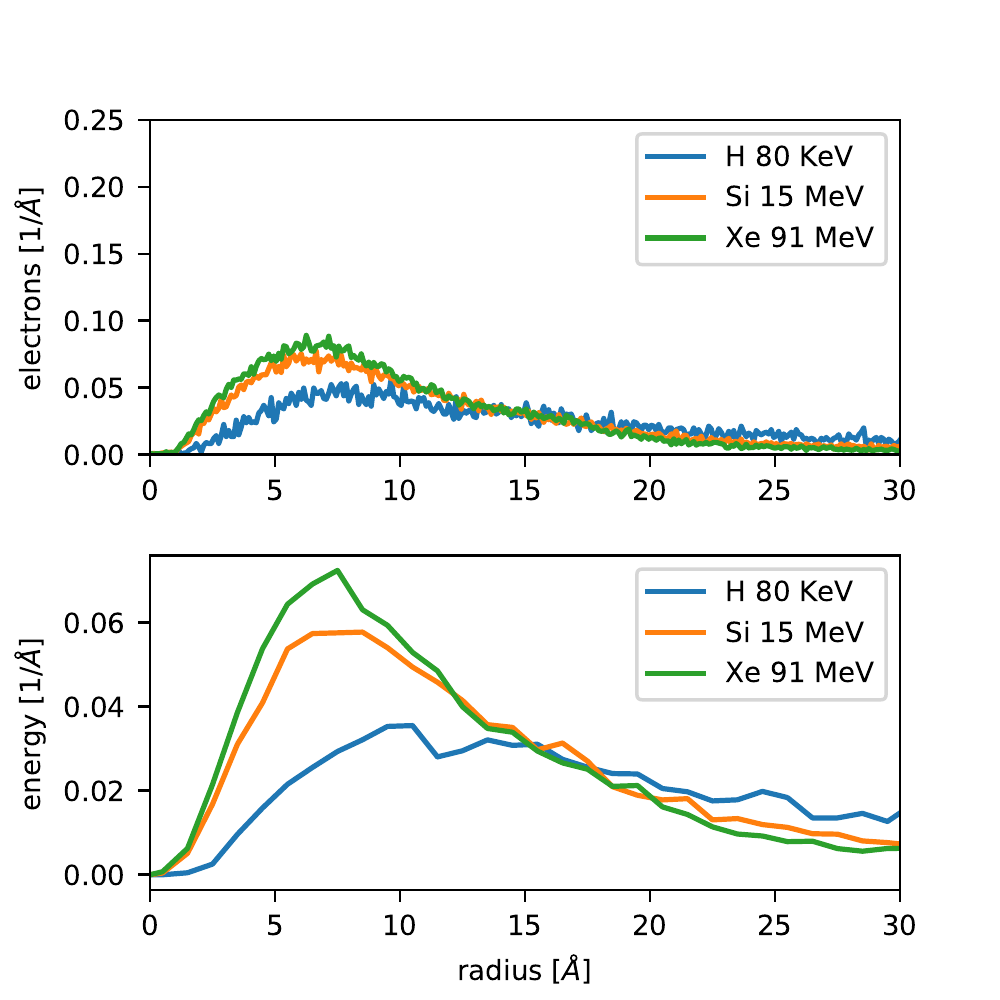}
    \caption{Distribution of the returned electron density (top graph) and summed kinetic energy of those electrons (bottom graph) in radial coordinates. The distribution is normalized to unity.  We show these results for the ions: H 80 keV (blue), Si 15 MeV (orange) and Xe 91 MeV (green).
    }
    \label{fig:radial_histogram}
\end{figure}

In figure \ref{fig:radial_histogram} we also show the radial distribution of the recaptured electron density by the layer, as well as the total kinetic energy carried by those electrons. We see that most of the electron density returns within 2 nm from the impact parameter. Similarly, the energy retrieved to the layer by the returning electrons is also concentrated within a radius of 3 nm. Among the ions tested, H 80 keV shows the most spread distribution. This is most likely due to the lower electrostatic potential generated by this ion.
\vspace{1em}

\begin{figure} [H]
    \centering
    \includegraphics[width=\linewidth]{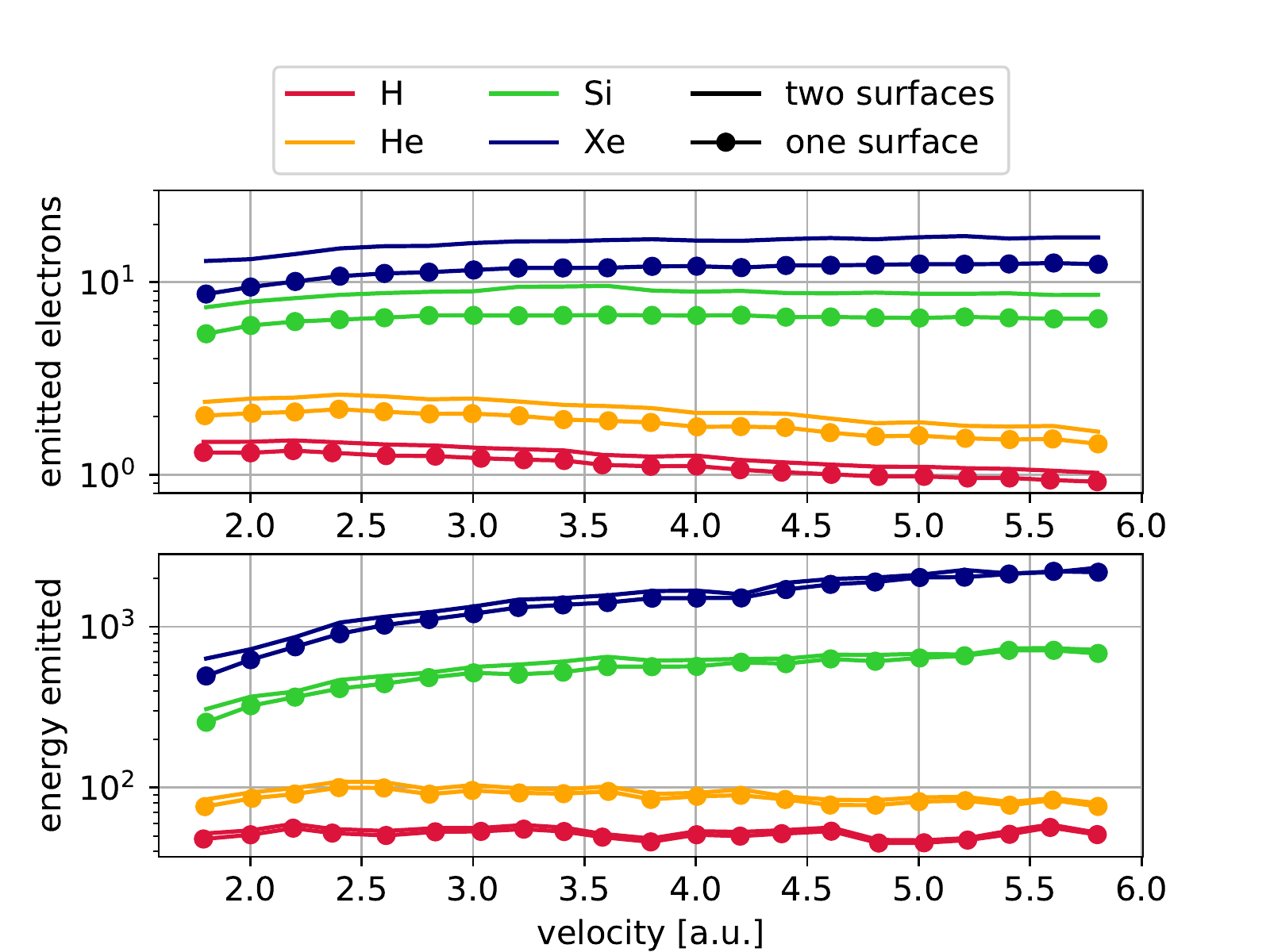}
    \caption{Comparison between PIC simulations of electron emission employing either one (line with points) or two (solid line) symmetric emission surfaces. The figures show the number of emitted electrons with velocity (top graph) and the energy removed via electron emission (bottom graph).  
    }
    \label{fig:one_vs_two_emission_surfaces}
\end{figure}

In figures \ref{fig:one_vs_two_emission_surfaces} we show the difference in electron emission and the energy emitted if one emission surface (only forward emission) or two (symmetric forward and backward emission) are assumed. We see that for heavier projectiles with higher stopping power such as Si and Xe, the difference between both modelling cases is noticeable, and the amount of emitted electrons is up to 30\% smaller when all the emission happens from a single emission surface. For lighter ions, instead we see that the choice of the emission surfaces has a small impact on the number of emitted electrons. In the lower figure we also show the resulting energy emitted from PIC simulations for both emission surface setups. We see that the difference between both modelling cases is very small; for the heaviest ion Xe, the difference is under 10\% and yet smaller for lighter ions.

\begin{figure} [H]
    \centering
    \includegraphics[width=\linewidth]{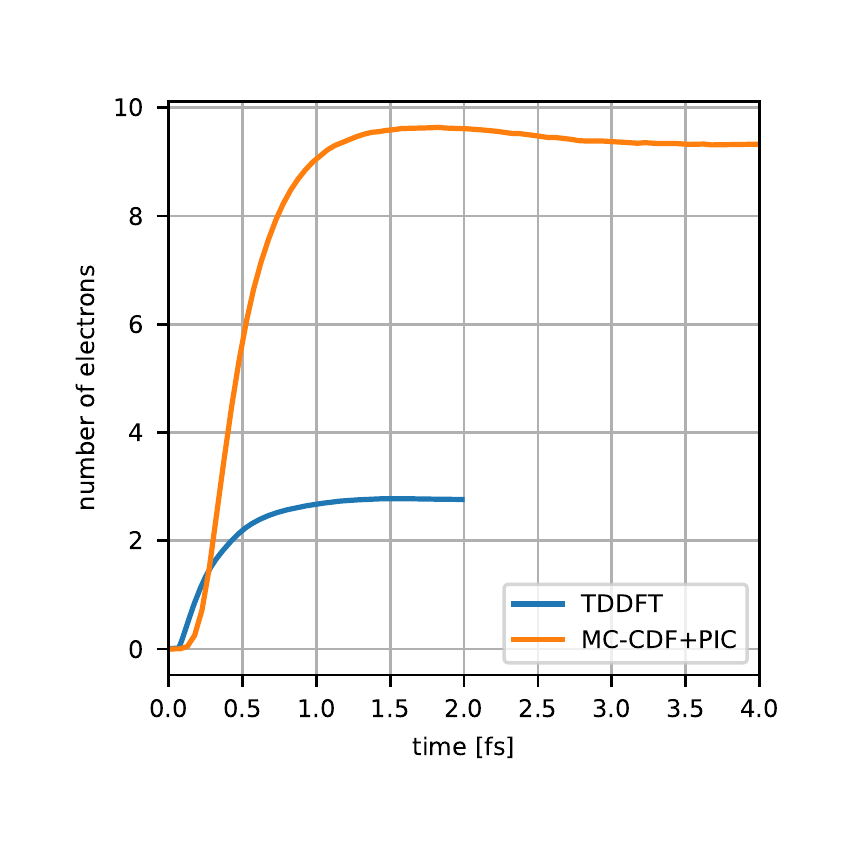}
    \caption{Evolution with time of the number of electrons outside the graphene layer in TDDFT (blue curve) and PIC (orange curve) for the ion Si$^{+4}$ 6 MeV. The electrons are considered in both methods as outside graphene when they are further than 10.5\,a$_{0}$ from the layer.}
    \label{fig:electrons_out_evolution}
\end{figure}

In figure \ref{fig:electrons_out_evolution} we show the evolution of the number of electrons outside graphene with time in PIC and TDDFT for the ion Si$^{+4}$ 6 MeV. We employ the same definition as in the main manuscript, and we consider the electrons to be outside the layer the electron density further than 10$a_{0}$ from the layer. We observe very similar dynamics in both methods, and within 1 fs the electron density outside the layer saturates. The total number of electrons outside the layer differs in both methods; this is, as we saw in the main text, is due to the different stopping power predictions in both approaches. Nevertheless, the similar dynamics clearly shows that the behaviour of the emitted electrons is captured similarly in both techniques.

\section{Additional MC-CDF results}
\subsection{Dependence on Density of States}
\begin{figure}
\includegraphics[width=0.95\columnwidth]{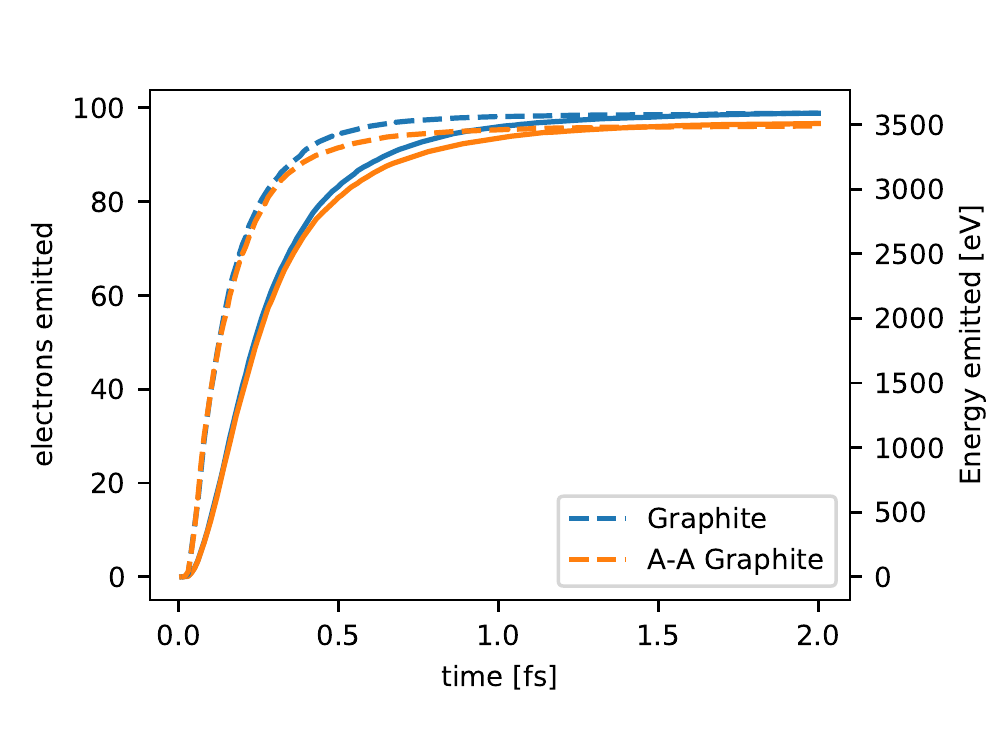}
\caption{\label{fig::DOS_comparison}
Number of electrons emitted (left axis) and energy emitted (right axis) in TREKIS as calculated with the experimental DOS \cite{ooi2006density} from graphite or with the A-A stacked graphite DOS obtained from DFT \cite{Vaz16}.
}
\end{figure}

The results shown in the manuscript were performed using the DOS from graphite\cite{ooi2006density} in the TREKIS code. The different electronic structure of graphene respect to graphite could lead to an additional systematic error in our calculations. To test this, we compared the cascade dynamics in TREKIS obtained employing the graphite DOS and using the A-A stacked DOS from previous work\cite{Vaz16}. We used for comparison the A-A stacked graphite DOS instead of graphene due to the fact that the calculations of the DOS of graphene include high energy vacuum states which are dependent on the cell size in the DFT calculations\cite{Vaz16}, whereas the bulk materials do not show that artifact. In figure \ref{fig::DOS_comparison} we show the electron emission as well as the percentage of energy emitted in TREKIS simulations during the impact of the ion Xe 91 MeV. We see that the DOS of graphite and A-A stacked graphite predict almost identical results. This shows that the use of the graphite DOS instead of the one of graphene barely affects the dynamics of electron emission investigated in this work.

\subsection{Charge state of projectiles}

\begin{figure} [H]
    \centering
    \includegraphics[width=\linewidth]{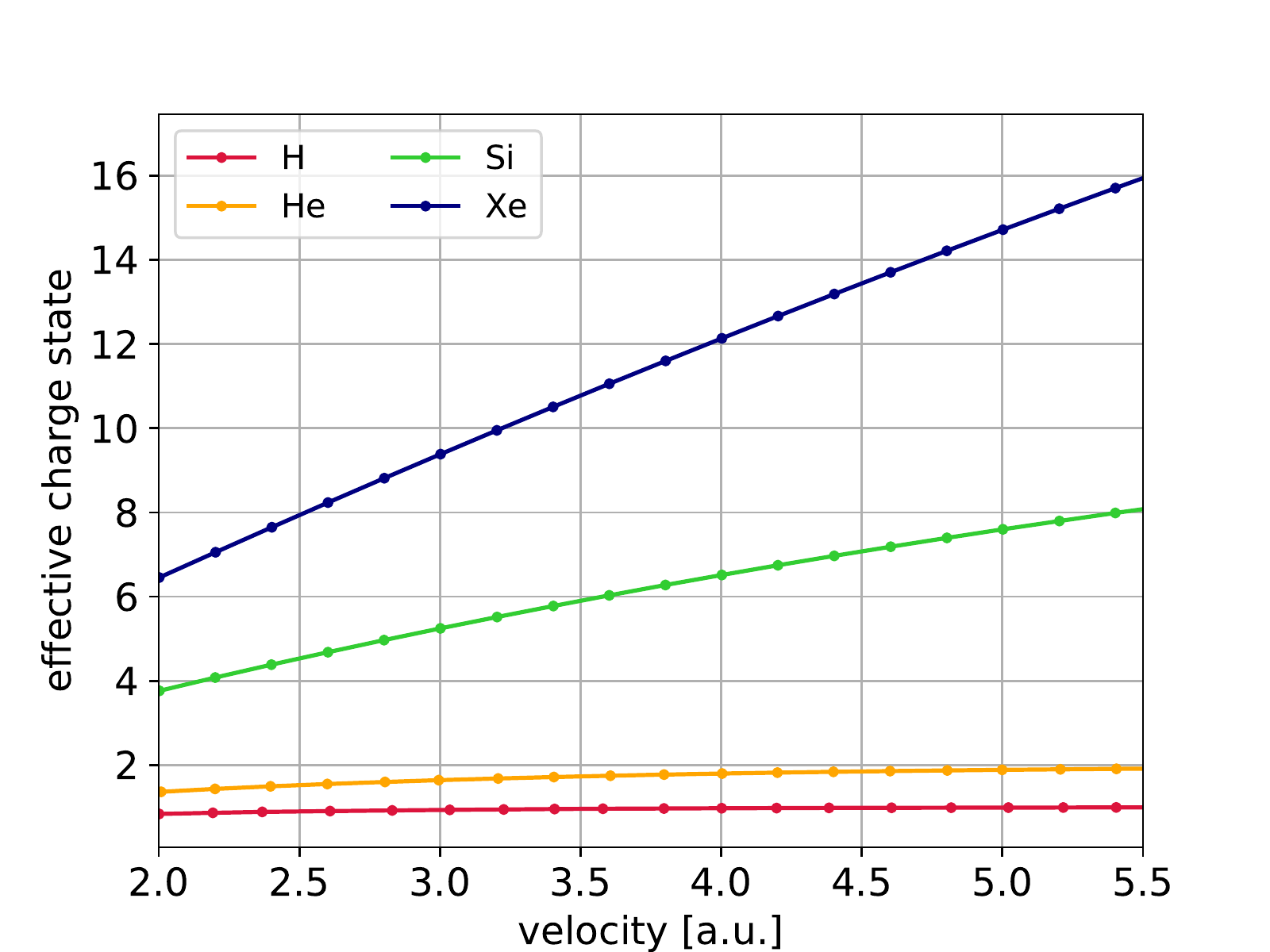}
    \caption{Effective charge state in MC-CDF for projectiles of different velocities. The charge states were computed according to the Barkas effective charge formula\cite{gervais1994simulation} for graphite.}
    \label{fig:charge_state}
\end{figure}

The MC-CDF calculations in the manuscript were performed assuming an equilibrium charge state for the ion. The effective charge of the projectile was described with the Barkas charge formula\cite{gervais1994simulation}
\begin{equation}
    q_{\mathrm{eff}} = Z(1-e^{-125 \beta Z^{-2/3}}),
    \label{eq:barkas}
\end{equation}
where $Z$ is the projectile atomic number and $\beta$ its relative velocity. This effective charge description was shown to give a good agreement in TREKIS with the data in the SRIM database\cite{medvedev2015time}. The effective charge values for each ion species can be found in Figure \ref{fig:charge_state}.

\subsection{Electron excitation in track core}
The MC-CDF method is a statistical approach, and therefore it does not  take into account the position of the atoms in the lattice; instead it assumes the atoms are distributed randomly in the material. Furthermore, this method does not consider the local availability of electrons and allows to excite more electrons than the total number of electrons from the nearby atoms. These features can potentially lead to an overestimation of the electron excitation in the model if the ionization in the material is large enough.

\begin{figure}
\begin{center}
\includegraphics[width=0.9\linewidth, trim={2cm 4cm 1cm 10cm}, clip]{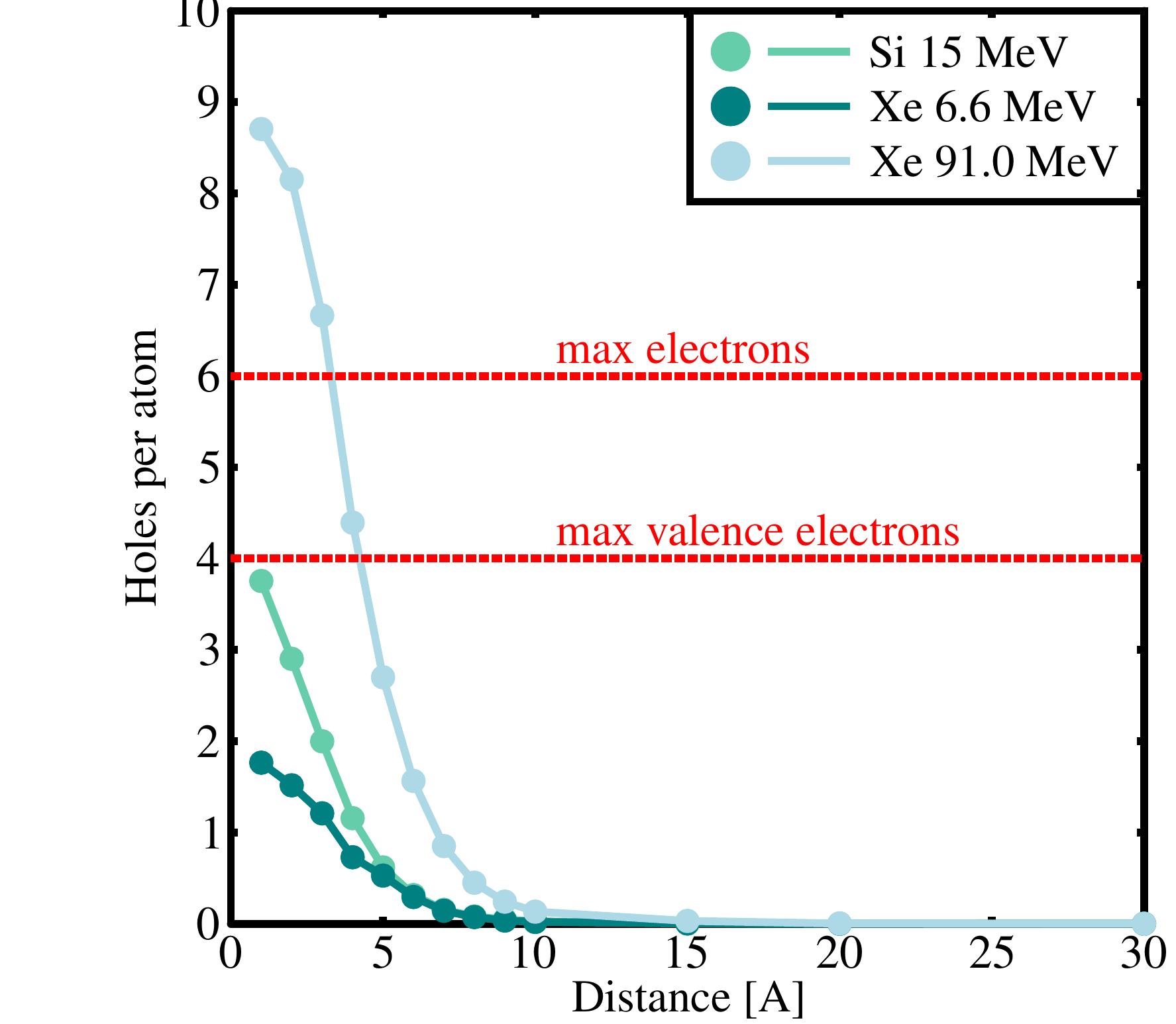}
\caption{Radial distribution of primary holes in MC for various ions at t=0.5 fs. The red lines show the number of valence electrons and total number of electrons available for excitation in a Carbon atom.}
\label{fig::holes_per_atom}
\end{center}
\end{figure}

We can estimate the density of primary electrons excited by the ion and see if on average in the track core there are more electrons excited than the number of valence electrons. In figure \ref{fig::holes_per_atom} we show the number of primary holes per atom along the radial direction excited by the projectile at 0.5 fs after the ion impact, we indicate the critical number of electrons which correspond to the full valence excitation (4 electrons per atom) and the full valence shell and core shell excitation (6 electrons per atom) with red lines. For most of the projectile energies used in this article, the number of holes never reaches the full ionization of the valence shells. Nevertheless, for heavy and faster ions such as Xe 91 MeV, the primary ionization can be particularly large, and the excitation may be heavily overestimated within few Angstrom radius of the impact point. By integrating the distribution we deduce that in the case of Xe 91 MeV the number of excited electrons in the layer exceed by 25\% the total number of electrons available for excitation in the region around the ion path.

\bibliographystyle{apsrev}
\bibliography{bibliography_CDF,bibliograhy_e_cascades,bibliography_tddft,bibliograhy1}